\documentclass[twocolumn,dvipsnames]{aastex62}

\usepackage{bm}
\usepackage{amsmath}

\newcommand\kms{\ifmmode{\rm km\thinspace s^{-1}}\else km\thinspace s$^{-1}$\fi}
\newcommand{\jykms}{\mathrm{Jy}\thinspace\mathrm{km}\thinspace \mathrm{s}^{-1}}

\newcommand{\vsgr}{V4046~Sgr}
\newcommand{\dq}{DQ~Tau}
\newcommand{\ak}{AK~Sco}
\newcommand{\uzt}{UZ~Tau}
\newcommand{\uzte}{\uzt~E}
\newcommand{\uztw}{\uzt~W}
\newcommand{\twelve}{${}^{12}$CO}
\newcommand{\thirteen}{${}^{13}$CO}
\newcommand{\eighteen}{C${}^{18}$O}
\newcommand{\figeight}{figure-$\mathsf{8}$}

\newcommand{\kepler}{\emph{Kepler}}

\newcommand{\imut}{\ensuremath{\theta}}

\begin{document}

\title{The Degree of Alignment Between Circumbinary Disks and Their Binary Hosts}

\correspondingauthor{Ian Czekala}
\email{iczekala@berkeley.edu}

\author[0000-0002-1483-8811]{Ian Czekala}
\altaffiliation{NASA Hubble Fellowship Program Sagan Fellow}
\affiliation{Department of Astronomy, University of California at Berkeley, Campbell Hall, CA 94720-3411, USA}

\author[0000-0002-6246-2310]{Eugene Chiang}
\affiliation{Department of Astronomy, University of California at Berkeley, Campbell Hall, CA 94720-3411, USA}
\affiliation{Department of Earth and Planetary Science, University of California at Berkeley, McCone Hall, Berkeley, CA 94720-4767, USA}

\author[0000-0003-2253-2270]{Sean M. Andrews}
\affiliation{Center for Astrophysics $\vert$ Harvard \& Smithsonian,
60 Garden Street, Cambridge, MA 02138, USA}

\author[0000-0002-4625-7333]{Eric L. N. Jensen}
\affiliation{Department of Physics and Astronomy,
Swarthmore College, 500 College Avenue, Swarthmore, PA 19081, USA}

\author[0000-0002-5286-0251]{Guillermo Torres}
\affiliation{Center for Astrophysics $\vert$ Harvard \& Smithsonian,
60 Garden Street, Cambridge, MA 02138, USA}

\author[0000-0003-1526-7587]{David J. Wilner}
\affiliation{Center for Astrophysics $\vert$ Harvard \& Smithsonian,
60 Garden Street, Cambridge, MA 02138, USA}

\author[0000-0002-3481-9052]{Keivan G. Stassun}
\affiliation{Department of Physics and Astronomy, Vanderbilt University, 6301 Stevenson Center, Nashville, TN 37235, USA}
\affiliation{Department of Physics, Fisk University, Nashville, TN 37208, USA}

\author[0000-0003-1212-7538]{Bruce Macintosh}
\affiliation{Kavli Institute for Particle Astrophysics and Cosmology,
Stanford University, 452 Lomita Mall, Stanford, CA 94305, USA}

\begin{abstract}
All four circumbinary (CB) protoplanetary disks orbiting short-period ($P < 20$\,day) double-lined spectroscopic binaries (SB2s)---a group that includes UZ~Tau~E, for which we present new ALMA data---exhibit sky-plane inclinations $i_{\rm disk}$ which match, to within a few degrees, the sky-plane inclinations $i_\star$ of their stellar hosts. Although for these systems the true mutual inclinations $\theta$ between disk and binary cannot be directly measured because relative nodal angles are unknown, the near-coincidence of $i_{\rm disk}$ and $i_\star$ suggests that $\theta$ is small for these most compact of systems. We confirm this hypothesis using a hierarchical Bayesian analysis, showing that 68\% of CB disks around short-period SB2s have $\theta < 3.0\degr$. Near co-planarity of CB disks implies near co-planarity of CB planets discovered by \kepler, which in turn implies that the occurrence rate of close-in CB planets is similar to that around single stars. By contrast, at longer periods ranging from 30--$10^5$ days (where the nodal degeneracy can be broken via, e.g., binary astrometry), CB disks exhibit a wide range of mutual inclinations, from co-planar to polar. Many of these long-period binaries are eccentric, as their component stars are too far separated to be tidally circularized. We discuss how theories of binary formation and disk-binary gravitational interactions can accommodate all these observations.
\end{abstract}

\section{Introduction} \label{sec:intro}
A fact no less true for being so commonly stated, most solar type stars do indeed reside in binary or higher multiplicity systems \citep{raghavan10}. The influence of binarity on star and planet formation has primarily been addressed from the perspective of circumstellar disks or planets perturbed by an external stellar companion \citep[i.e., in the ``S''-type configuration;][]{dvorak82}. For example, the presence of a binary companion with semi-major axis $a < 50$\,au will truncate the outer radius of a circumstellar disk \citep{jensen96a,harris12}, possibly reducing the planet occurrence rate \citep{wang14,kraus16}. By comparison, at the closest binary separations ($a \lesssim 10$\,au), massive disks will most likely be circumbinary \citep[i.e., in the ``P''-type configuration;][]{dvorak82,harris12}.

Efforts to study planet formation in the circumbinary sense were given new urgency by the \kepler\ mission's discovery of circumbinary (CB) planets orbiting eclipsing binary (EB) stars \citep[e.g.,][]{doyle11,welsh12,orosz12,schwamb13}. Thus far, the dozen known transiting CB planets orbit binaries whose periods $P$ fall in the range of 7 to 40 days and are inclined relative to their host EB planes by $\imut < 5\degr$ \citep[see the compilation by][]{winn15}. The underlying distribution of mutual inclinations is not well constrained. Because the \kepler\ CB planet sample is sourced from the \kepler\ EB sample (which have sky-plane inclinations $i_\ast \simeq 90\degr$; i.e., the binaries are viewed nearly edge-on), the \kepler\ CB planet survey has poor sensitivity to planets on orbits with large mutual inclinations. Even if an inclined planet were to transit once, subsequent transits occurring every planetary orbital period would not be guaranteed because the transited star would have moved in its orbit \citep{martin14}. Knowing the mutual inclination distribution is necessary for calculating the intrinsic occurrence rate of \kepler\ CB planets \citep{armstrong14}. Statistical and dynamical arguments suggest $\theta \lesssim 3\degr$ \citep{li16}, while the non-detections of the BEBOP radial velocity survey, which is sensitive to CB giant planets around single-lined EBs with periods $P < 40$\,days, can be combined with \kepler\ statistics to infer that $\theta \lesssim 10\degr$ \citep{martin19}. Other detection techniques such as radial velocity, eclipse timing variations, microlensing, and direct imaging have also been employed to search for CB planets, though their samples are smaller than \kepler's and their survey selection functions are often more difficult to characterize \citep[e.g.,][]{udry02,bailey14,bennett16,asensio-torres18,derosa19}. 

\begin{figure}[tb]
\begin{center}
  \includegraphics{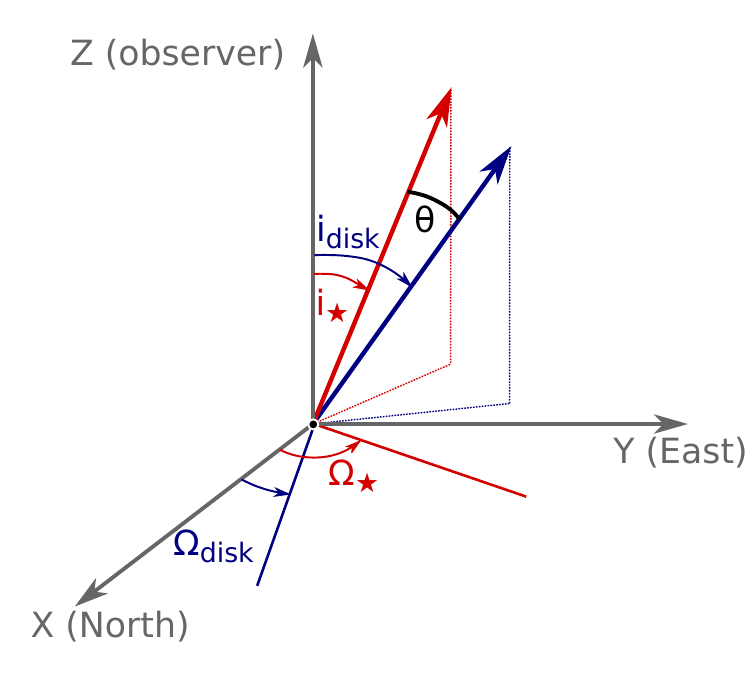}
  \figcaption{The orientations of a binary orbit and its circumbinary disk are denoted using their unit orbital angular momentum vectors (in red and blue, respectively). The mutual inclination $\imut$ is the angle between these two vectors. For an observer located at $Z=+\infty$, the inclination $i$ is defined as the angle between the orbit vector and the $Z$ axis. The position angle of the ascending node $\Omega$ defines the orientation of the orbit in the tangent sky-plane (the $X$-$Y$ plane).
  \label{fig:coords}}
  \end{center}
\end{figure}

Another way to get at CB planet inclinations, and gain broader insight into the formation of CB planets in general, is to study CB disks, either of the protoplanetary or debris variety. Spatially-resolved observations of CB disks constrain their  sky-plane inclinations $i_\mathrm{disk}$ and position angles of their ascending nodes $\Omega_\mathrm{disk}$ (see Figure~\ref{fig:coords}).\footnote{We follow the convention of the visual binary field and throughout this paper label the ascending node as the one at which the secondary is \emph{receding} from the observer \citep[e.g.,][]{vandekamp81,torres95,pourbaix98}, whose position angle $\Omega$ is measured by the number of degrees east of north (counter clockwise on the sky). The approaching node is sometimes used as the ascending node in the exoplanet field \citep[e.g.,][]{murray10}.} Combining these disk parameters with measurements of the sky-plane binary inclination $i_\star$ and ascending node $\Omega_\star$  enables calculation of the mutual disk-binary inclination $\imut$: 
\begin{equation}
    \cos \imut = \cos i_\mathrm{disk} \cos i_\star + \sin i_\mathrm{disk} \sin i_\star \cos(\Omega_\mathrm{disk} - \Omega_\star).
    \label{eqn:fekel}
\end{equation}
Joint radial velocity (RV) and astrometric observations yield $i_\star$ and $\Omega_\star$ directly. For the shortest period binaries that are astrometrically inaccessible, it is still possible to measure $i_\star$ by combining a double-lined spectroscopic radial velocity measurement of $(M_1 + M_2)\sin^3 i_\star$ with an estimate of the total stellar mass $M_1 + M_2$ from the disk rotation curve. This leaves only $\Omega_\star$ unspecified. Unfortunately, with a uniform prior on $\Omega_\star$, there still remains a wide range of possible mutual inclinations, particularly for systems not viewed face on.

The situation for short-period systems, however, is not necessarily as uncertain as the above considerations suggest. Suppose the intrinsic distribution of disk-binary inclinations were broad. In that case, it would be surprising to measure $i_{\rm disk}$ and $i_\star$ in a given system and find these two angles to be nearly identical. The more often we observed this near-equality, the more surprised we would be. And yet that is precisely the hand we have been dealt: of the four known CB protoplanetary disks orbiting $P < 20$ day double-lined spectroscopic binaries (SB2s), all have $i_{\rm disk} \simeq i_\star$ to within a few degrees. This measurement outcome leads us to suspect that the deck is stacked: that we do not live in a universe where mutual inclinations are random, but one in which they are preferentially small, at least for these most compact of systems.

A large portion of this paper is devoted to proving, in a statistically rigorous manner, that most circumbinary disks orbiting short-period binaries are nearly co-planar (see also \citealt{prato02}, \citealt{jensen07}, and \citealt{kennedy12a}). We will implement a hierarchical Bayesian analysis that leverages the (incomplete) data we have for CB disks orbiting short-period SB2s to infer, in full, the intrinsic distribution of mutual inclinations $\theta$ from which they are drawn. We will supplement this analysis by compiling a database of CB disks orbiting longer period binaries, to search for possible trends between disk-binary mutual inclination and other system parameters such as binary period and binary eccentricity.

We begin our study by using new Atacama Large Millimeter/Submillimeter Array (ALMA) data to update the parameters of one of the four CB disks orbiting short-period SB2s, \uzte. In \S\ref{sec:data}, we present \twelve, \thirteen, and \eighteen\ $J=2-1$ data for \uzte, derive a new dynamical mass for the central binary, and measure $i_\star$ (but not $\Omega_\star$). In \S\ref{sec:sample} we assemble a collection of circumbinary protoplanetary and debris disk systems from the literature and calculate their mutual inclinations---in some cases only in a statistical sense. There we implement a hierarchical Bayesian model to infer the underlying mutual inclinations of the subsample of CB disks around SB2s. In \S\ref{sec:discussion} we discuss our observations in the context of theories of binary formation and disk-binary gravitational interactions, and make connections to the population of \kepler\ CB planets. We conclude in \S\ref{sec:conclusion}.

\section{UZ Tau Data and Analysis} \label{sec:data}
We review what is known about \uzt, which actually comprises two binaries, in \S\ref{subsec:background}. We describe our new ALMA data and how we reduced it in \S\ref{subsec:new_alma}. By forward-modelling the molecular line emission, we calculate disk and binary inclination parameters for \uzte\ in \S\ref{subsec:method}.

\subsection{Background data on \uzt} \label{subsec:background}
The \uzt\ system consists of \uzte, which is a double-lined spectroscopic binary consisting of Ea and Eb \citep[$P = 19.131$\, days, $e = 0.33$, $q \equiv M_2/M_1 = 0.30$;][]{mathieu96,jensen96b,prato02,jensen07}, and \uztw, which is a visual binary consisting of Wa and Wb, separated by 0\farcs3. The E and W binaries are separated by 3\farcs6 \citep{correia06}. The Ea and Eb stars have spectral types M1 and M4, respectively \citep{prato02}, and the Wa and Wb stars are both M2 spectral type \citep{correia06}. A \emph{Gaia} DR2 parallax exists for \uzte\ of $\pi=7.62\pm 0.10\,$mas \citep[including a $0.02\,$mas systematic term,][]{lindegren18} or $131.2 \pm 1.7\,$pc  \citep{gaia18,bailer-jones18}.

\citet{simon00} observed the \uzte\ CB disk using \twelve\ $J=2-1$ observations from the IRAM interferometer. They forward-modeled the disk rotation to derive a total stellar mass of $M_\mathrm{Ea} + M_\mathrm{Eb} = 1.31 \pm 0.08\,M_\odot$ (assuming $d = 140\,\mathrm{pc}$).  However, some uncertainty remains in this result due to severe \twelve\ contamination from the molecular cloud and the modest spatial and spectral resolution of the observations (1\farcs1 and 1.2\,\kms, respectively). Nevertheless, \citet{prato02} combined this disk-based measurement with the double-lined radial velocity solution to find $|i_\mathrm{disk} - i_\star| < 5\degr$. We obtained new high resolution ALMA observations of the \uzt\ system in the \twelve, \thirteen, and \eighteen\ $J=2-1$ transitions in order to derive new dynamical mass measurements of \uzte\ and remove the lingering uncertainties.

Recent sub-mm, mm, and cm-wave continuum observations of \uzt\ have also clarified the distribution of dust in the quadruple system. \citet{tripathi18} studied \uzt\ across a broad range of radio and millimeter wavelengths and spatially resolved individual circumstellar disks around Wa and Wb for the first time. \citet{long18} targeted \uzte\ with high resolution (0\farcs12) ALMA observations of the dust continuum. They found disk substructures in the form of a small inner cavity ($r < 15\,$au) and two low-contrast axisymmetric emission rings at $\sim$$20\,$au and $\sim$$75\,$au. They also measured the disk inclination (relative to the sky plane) to be $i_\mathrm{disk} = 56.15 \degr \pm 1.50\degr$ and the position angle of the ascending node to be $\Omega_\mathrm{disk} = 90.39 \pm 1.50\degr$ (systematic uncertainties included).

\subsection{New ALMA observations of \uzt} \label{subsec:new_alma}
We observed the \uzt\ system with ALMA on June 21, 2016 under project code 2015.1.00690.S. The array was comprised of 36 antennas and the baselines ranged from 15 - 704\,m. This configuration of the array yielded a maximum recoverable scale (MRS) of $10.7\arcsec$. The total on-source time was 22 minutes. Our band 6 observations targeted the dust continuum with a 1.875\,GHz wide spectral window centered at 233.000\,GHz, and we placed three spectral windows on the \twelve, \thirteen, and \eighteen\ $J$=2-1 molecular transitions with channel spacings of 30.5, 30.5, and 61.0\,kHz, respectively. The Hanning smoothing window applied by the ALMA correlator slightly correlates the information content of adjacent channels, suppressing Gibbs ringing but effectively reducing the resolution by an additional factor of $\approx$2, resulting in effective velocity resolutions of 79, 83, and $167\,\rm m\thinspace s^{-1}$, respectively.

To create a set of continuum-only visibilities, we used the CASA task \texttt{plotms} \citep{mcmullin07} to identify channels with strong line emission, excised these, and averaged the remaining channels to create a measurement set with a total of 2.344\,GHz continuum bandwidth. We used the CASA task \texttt{clean} to image the continuum and placed masks around the visible sources \uzt\ E and W. We used this CLEAN model to perform three rounds of phase-only self-calibration and one round of amplitude and phase self-calibration via the \texttt{gaincal} task. After applying the solution to the measurement set, the final RMS in the continuum channel was reduced from 450\,$\mu$Jy\,$\mathrm{beam}^{-1}$ to 70\,$\mu$Jy\,$\mathrm{beam}^{-1}$.

We then applied the self-calibration solution to the full dataset, including the spectral windows containing line emission. The continuum was estimated in the visibility plane and subtracted from the dataset using the CASA task \texttt{uvcontsub}. An initial round of imaging with \texttt{tclean} revealed that emission from all isotopologues was clearly detected, with the brightest, \twelve, spanning a linewidth of nearly 11\,\kms. We used the task \texttt{mstransform} to correct the spectral line channels to the reference frame of the kinematic local standard of rest (LSRK), and binned all line spectral windows into matching channels $334\,\rm m\thinspace s^{-1}$ wide. We chose this channel width, representing binning factors of $\sim$8, $\sim$8, and 4 for the \twelve, \thirteen, and \eighteen\ spectral windows, respectively, because this produces effectively independent channels for all transitions on the same velocity scale and is a sensible compromise between spectral resolution and anticipated computational burden, which scales with the number of channels. While in principle the line emission could be analyzed at higher spectral resolution with less aggressive binning factors, the considerable linewidth of \uzte\ means that the disk rotation curve is still sampled with more than 30 effective channels at the chosen resolution.

\begin{figure*}[ht!]
\begin{center}
  \includegraphics{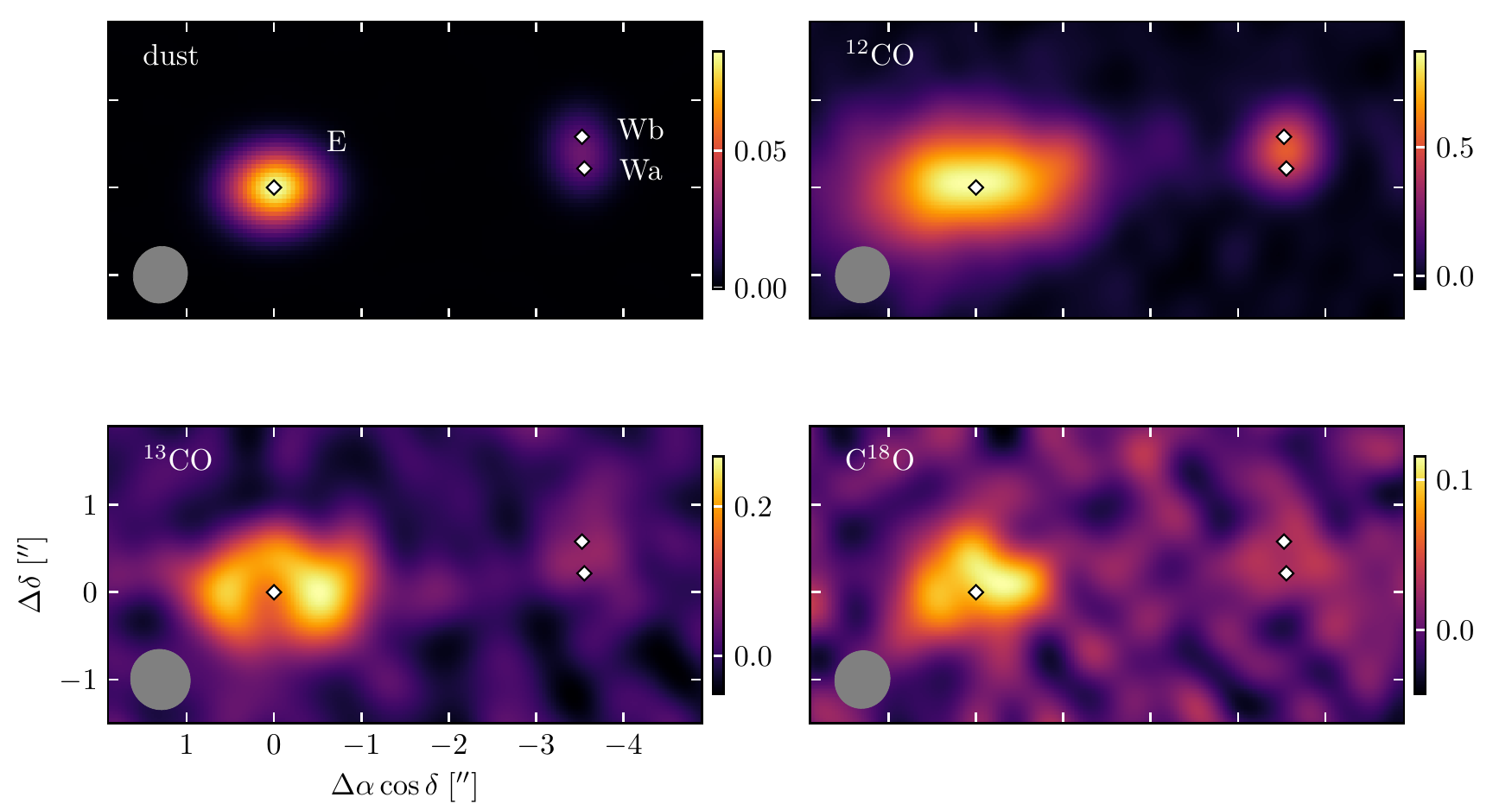}
  \figcaption{The 233\,GHz continuum image and \twelve, \thirteen, and \eighteen\ moment-$\mathsf{0}$ maps (velocity-integrated intensity). The units of the continuum image are Jy\,$\mathrm{beam}^{-1}$, while the moment-$\mathsf{0}$ maps are $\jykms\,\mathrm{beam}^{-1}$. The synthesized beam geometry is shown in the lower left corner of each plot. The maps are centered on the \uzte\ continuum emission and overlaid with the near-infrared positions of the \uzt\ Wa and Wb stars relative to the \uzte\ spectroscopic binary \citep{correia06}.
  \label{fig:moments}}
  \end{center}
\end{figure*}

We produced a final image of the dust continuum of \uzt\ E and W using the multiscale-multifrequency \texttt{clean} algorithm and Briggs parameter \texttt{robust=0.5} (see Figure~\ref{fig:moments}). We fit the dust emission of \uzte\ with an elliptical Gaussian using the \texttt{uvmodelfit} task, and find a flux density of $145.72 \pm 0.04$\,mJy (the uncertainty does not include systematic calibration uncertainties). The dust emission from the western component is extended along a north-south axis. Recently, \citet{tripathi18} published Karl G. Jansky Very Large Array (JVLA) observations at 30.5\,GHz and 37.5\,GHz with 0\farcs1 resolution, and individually resolved the Wa and Wb circumstellar disks, showing that at these frequencies, they are near equal-brightness.
Although we are unable to individually resolve the Wa and Wb components with the $0\farcs6$ beam of our observations, the elongated emission is indicative that we are seeing emission coming from both the Wa and Wb components. The total measured flux from these is 34\,mJy, consistent with the 225\,GHz Combined Array for Research in Millimeter-wave Astronomy (CARMA) measurements also presented in \citet{tripathi18}.

\begin{deluxetable}{lcc}[tb]
\tablewidth{18pc}
\tablecaption{ALMA Image Properties  \label{tab:ims}}
\tablehead{
\colhead{} &
\colhead{beam dimensions, P.A.} &
\colhead{RMS [mJy beam$^{-1}$]}}
\startdata
233\,GHz cont.  & $0\farcs66\times0\farcs61$, -25\degr\ & 0.070 \\
$^{12}$CO $J$=2$-$1 & $0\farcs65\times0\farcs61$, -26\degr\ & 7     \\
$^{13}$CO $J$=2$-$1 & $0\farcs70\times0\farcs68$, 35\degr\ & 8     \\
C$^{18}$O $J$=2$-$1 & $0\farcs66\times0\farcs63$, -19\degr\ & 5     \\
\enddata
\tablecomments{The RMS noise levels recorded for the spectral line cubes correspond to the values per 334\,m s$^{-1}$ channel. All images were synthesized with \texttt{robust=0.5}.}
\end{deluxetable}

We synthesized images of the CO isotopologue line emission using the CASA \texttt{tclean} task with Briggs \texttt{robust=0.5} weighting. We used the \texttt{auto-multithresh} to generate an initial starting mask for each CO isotopologue, which was then edited by hand channel-by-channel to conform to the observed emission from the E and W components. The properties of the synthesized images are summarized in Table~\ref{tab:ims}, and the channel maps are shown in the Appendix (Figures~\ref{fig:chmaps-12}, \ref{fig:chmaps-13-18}, \& \ref{fig:chmaps-high}).

Spatially-resolved channel maps can be useful for determining the systemic velocity of a protoplanetary disk because they generally exhibit reflective symmetry across the disk minor axis, about the systemic velocity. The central channel at the systemic velocity contains two lobes in a characteristic ``figure $\mathsf{8}$'' shape, while lobes in  redshifted or blueshifted channels stretch into ``$\mathsf{C}$-like'' shapes. Based upon visual inspection of the \twelve\ maps synthesized at the highest $79\,\rm m\thinspace s^{-1}$ native resolution (Appendix, Figure~\ref{fig:chmaps-high}), it is clear that the \uzte\ ``figure $\mathsf{8}$'' is located at $v_\mathrm{LSRK} = 5.5 \pm 0.1\,\kms$ ($v_\mathrm{bary} = 15.5 \pm 0.1\,\kms$).\footnote{In the direction of \uzt, the conversion between the kinematic local standard of rest (LSRK) and the barycentric velocity is $v_\mathrm{bary} = v_\mathrm{LSRK} + 10.00\,\kms$.} The \twelve\ channel maps in \citet{simon00} appear consistent with our determination of the systemic velocity---their coarsely spaced channel maps ($1.2\,\kms$) show the ``figure $\mathsf{8}$'' at or near $v_\mathrm{LSRK} = 5.4\,\kms$. By forward-modeling single-dish CN observations, \citet{guilloteau13} obtained a systemic velocity of $v_\mathrm{LSRK} = 5.96\pm0.47\,\kms$, which is consistent with our determination given the large uncertainty on their measurement. \citet{guilloteau13} also determined a systemic velocity by analyzing the wings of \thirteen\ emission, which they argued should be minimally affected by cloud contamination, and obtained $v_\mathrm{LSRK} = 5.90 \pm 0.18\,\kms$. It is unclear how cloud-contaminated channels were identified and masked in this analysis, and so it is possible this systemic velocity measurement may have been affected by residual cloud contamination. Our systemic velocity is mildly discrepant ($2\sigma$) from that determined spectroscopically \citep[$\gamma_\mathrm{LSRK} = 3.9\pm0.7\,\kms$,][]{jensen07}. This difference likely originates from uncalibrated RV zeropoint offsets, which are frequently of this magnitude and spectrograph-dependent.

Deviations from the reflective symmetry of CO channel maps---most visible as ``missing'' flux---indicate that the \uzt\ line emission is contaminated by a foreground molecular cloud. We generated moment-$\mathsf{0}$ maps (total intensity integrated over all velocity channels) for the \twelve, \thirteen, and \eighteen\ isotopologues using the \texttt{immoments} CASA task summing across all spectral channels in the range $0.0 - 13.0\,\kms$ LSRK. We detect faint line emission from \uztw\ in all CO isotopologues (see Figure~\ref{fig:moments}). Using the moment-$\mathsf{0}$ maps for each isotopologue as a guide, we created new masks encapsulating the full spatial extent of the emission, for each transition, for both \uzt\ E and W. In contrast to the CLEAN masks used to synthesize the channel maps, which have different shapes for each channel, these masks were the same shape for all channels. We calculated the total integrated flux for the lines using these masks and the \texttt{imstat} task on the moment-$\mathsf{0}$ maps. For \uzte, the \twelve, \thirteen, and \eighteen\ line fluxes are $6.01\pm0.02\,\jykms$, $0.99\pm0.02\,\jykms$, and $0.35\pm0.01\,\jykms$, respectively. For \uztw, the line fluxes are $0.77\pm0.02\,\jykms$, $0.13\pm0.02\,\jykms$, and $0.06\pm0.01\,\jykms$, respectively. We note that because of the foreground cloud contamination, these line fluxes are lower limits only.

To confidently identify which specific channels suffer cloud contamination and excise them from the dynamical analysis, we inspected the \twelve\ and \thirteen\ $J$=1-0 maps of the Five College Radio Astronomy Observatory CO Mapping Survey of the Taurus Molecular Cloud \citep[FCRAO;][]{goldsmith08,narayanan08}. This survey was carried out with the 13.7\,m single-dish Quabbin telescope and the maps have a spatial resolution of 45\arcsec\ and velocity resolution of $1\,\kms$. At the location of the \uzt\ system, we find that there is significant \twelve\ and \thirteen\ cloud emission for channels in the velocity range $v_\mathrm{LSRK} = 4 - 7\,\kms$, and a faint trace of emission in the $7-8\,\kms$ channel. Therefore, we choose to mask from our analysis all channels in the range $4.0 \leq v_\mathrm{LSRK} \leq 7.5\,\kms$.

\begin{figure}[tb]
\begin{center}
  \includegraphics{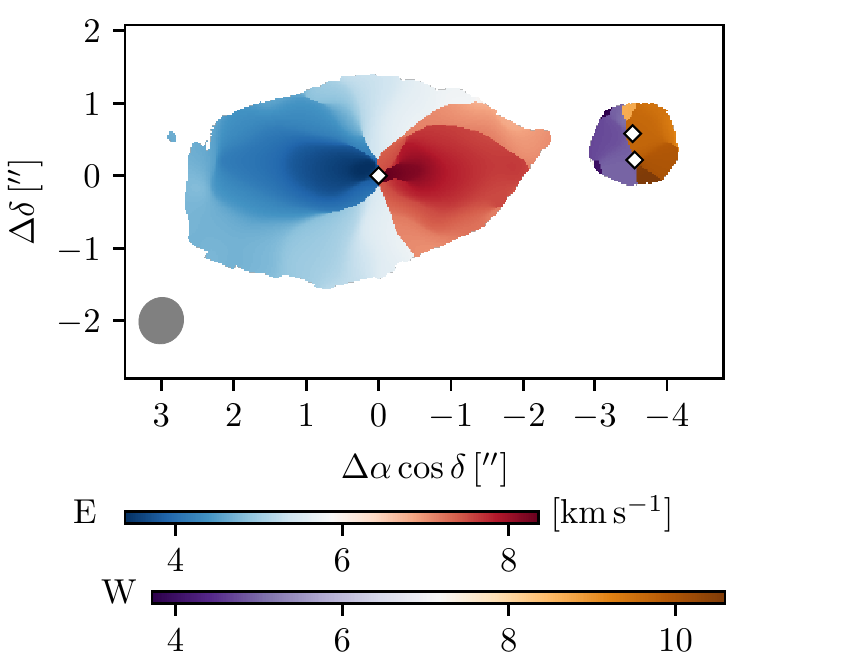}
  \figcaption{The velocity field of the \uzt\ system, generated from the \twelve\ emission using a quadratic fit as implemented in \citet{teague18} and masking all $\mathsf{0}$-th moment pixels below the 5th intensity percentile. The colorbars indicate the inferred velocity at each pixel for both \uzte\ and \uztw, and are stretched independently. We caution that due to cloud contamination, the retrieved velocities may be partially biased in the spatial pixels containing emission at $v_\mathrm{LSRK} = 4.0 - 7.5\,\kms$ (see Figure~\ref{fig:chmaps-12}). The western emission is consistent with originating from two small circumstellar disks around \uzt\ Wa and Wb, as seen in $\sim$30\,GHz dust emission by \citet{tripathi18}.
  \label{fig:moment-1}}
  \end{center}
\end{figure}

Using the quadratic-fit technique implemented by \citet{teague18} we made maps of the velocity field of the \uzt\ system as probed by \twelve\ emission (see Figure~\ref{fig:moment-1}). The velocity field of the combined emission of \uztw\ has a nodal position angle $\Omega_\mathrm{disk}$ that approximately matches that of \uzte. Given the \citet{tripathi18} observations, we know that Wa and Wb host individual circumstellar dust disks. The emission we see is consistent with both of these stars hosting gas disks. Higher spatial resolution observations in \twelve\ or other dense-gas tracers would definitively associate the western gas emission with Wa and/or Wb, measure the inclination of the Wa and Wb disks, and permit a measurement of coplanarity between the E and Wa and Wb disks. 

Notwithstanding the potential biases from cloud contamination, it is still worthwhile to compare the velocity extent of the \uzte\ and W emission. From Figure~\ref{fig:moment-1}, it is clear that one or both of the W disks host emission at larger radial velocities than the E disk. The linewidth of the emission is relatively large considering that the Wa and Wb stars are less than half of the mass of the E spectroscopic binary. This would seem to indicate that either the Wa and Wb disks are observed at higher inclination, their stellar host masses are greater than their spectral type would seem to imply, or both.

\citet{hartmann86} used optical spectroscopic observations of W to determine a systemic velocity of $v_\mathrm{LSRK} = 8.5\,\kms$ ($v_\mathrm{bary} = 18.5\,\kms$); however, in that analysis, the binary nature of W was not yet known. Taken at face value, given the large separation on the sky between the E and W systems, the $\Delta v_r \approx 3\,\kms$ radial velocity discrepancy between E and W would be suggestive that the two binary systems are not gravitationally bound to each other. However, given the proximity of E and W and the considerable uncertainty in the radial velocity of W (E. L. N. Jensen, \emph{private communication}), the possibility remains that the systems are bound. 

\subsection{UZ Tau E ${}^{13}\mathrm{CO}$ and $\mathrm{C}^{18}\mathrm{O}$ dynamical analysis} \label{subsec:method}
To determine the total mass $M_\mathrm{tot} = M_\mathrm{Ea} + M_\mathrm{Eb}$ of the \uzte\ spectroscopic binary, we model the rotation of the circumbinary disk as traced by molecular line emission. Previously, we noted that the \uzt\ system suffers from cloud contamination. Because \twelve\ is the isotopologue most readily affected by cloud contamination, we take a conservative approach and only model the emission from the less abundant \thirteen\ and \eighteen\ isotopologues in channels outside of the contaminated velocity ranges ($v_\mathrm{LSRK} = 4.0 - 7.5\,\kms$).

Briefly, our dynamical mass technique works by forward-modeling the molecular emission directly to the visibility plane. The disk structure is set by a temperature power law (with exponent $q$ and normalization at 10\,au by $T_{10}$) and a surface density power law (exponent $\gamma = 1$) with an exponential taper outside of the disk characteristic radius, $r_c$. The density power law is normalized by the total disk mass ($M_\mathrm{disk}$). For a given radius, the disk is assumed to be vertically isothermal and the vertical density distribution set by hydrostatic equilibrium. The velocity field of the disk is assumed to be Keplerian, and entirely determined by the central stellar mass ($M_\mathrm{tot}$). We assume the distance to \uzte\ to be $d = 131.2$\,pc. The disk structure is oriented relative to the observer based upon geometric parameters (inclination relative to the sky tangent-plane $i_\mathrm{disk}$, position angle of the ascending node $\Omega_\mathrm{disk}$, and offsets from the phase center $\delta_\alpha$ and $\delta_\delta$), and then channel maps are ray-traced using the \texttt{RADMC-3D} software package \citep{dullemond12}. The channel maps are then Fourier transformed, sampled at the baselines corresponding to the array position during the observations, and compared to the visibilities with a complex-valued $\chi^2$ likelihood function. After incorporating uniform priors on the disk structure parameters and a geometrical prior on the disk inclination, the posterior distribution is sampled using our \texttt{Julia} implementation of the Markov Chain Monte Carlo (MCMC) affine-invariant ensemble sampler \citep{goodman10,foreman-mackey13}.
For more information on the implementation of these procedures see \citet{czekala15a} and the \texttt{DiskJockey} codebase.\footnote{\url{https://github.com/iancze/DiskJockey}}

Most molecular line observations of disks leave a degeneracy as to whether $i_\mathrm{disk} < 90\degr$ or $i_\mathrm{disk} > 90\degr$ (orbits moving counter-clockwise or clockwise on the sky-plane, respectively). Although gas observations reveal instantaneous line-of-sight velocity information throughout the entire disk, they do not provide the sense of disk rotation in the same manner as astrometric binary orbits. The inclination degeneracy can sometimes be broken with very sensitive molecular observations, since the brighter of the two lobes of the \figeight\ at the systemic velocity indicates the far side of the disk \citep{rosenfeld13a}. For \uzte, the brightness asymmetry seen in the high resolution \twelve\ channel maps (Figure~\ref{fig:chmaps-high}) indicates that $i_\mathrm{disk} < 90^\circ$. If there are spatially resolved scattered light images of the disk, the strong forward-scattering of micron-sized dust grains can also indicate the near side of the disk. Since the high resolution observations by \citet{long18} of the dust continuum yield a more precise constraint on the disk inclination ($i_\mathrm{disk} = 56.15 \pm 1.5\degr$) than can be expected from our lower resolution data, we fix the disk inclination to this value in order to reduce the parameter space of our model and speed convergence of the MCMC ensemble.

We experimented with various temperature profiles. Our first experiments letting the normalization and exponent float yielded flat ($q \approx 0$) and cold ($T_{10} \approx 12$\,K) temperature profiles for both \thirteen\ and \eighteen. We then tried two temperature profiles with fixed power law exponents of $q=0.50$ and $q=0.75$, which are common values for flared and geometrically thin-disks, respectively. The choice of $q=0.50$ yielded the most reasonable temperature normalization ($T_{10} \approx 50$\,K) and a temperature profile consistent with the dust temperature profile presented in \citet{long18}. The full constraints on the model parameters using the \thirteen\ and \eighteen\ transitions are listed in Table~\ref{table:components}. From the \thirteen\ and \eighteen\ isotopologues, the total stellar mass is inferred to be $M_\mathrm{tot} = 1.19 \pm 0.01\,M_\odot$ and $M_\mathrm{tot} = 1.23 \pm 0.01\,M_\odot$, respectively (statistical errors only). Because these measurements are made with independent datasets but are statistically inconsistent with each other, this indicates that there is at least a $0.02\,M_\odot$ systematic error affecting these results, which is probably a result from missing complexity in our disk structure model (e.g., a vertical temperature gradient, CO freezeout). We average these two results and add the uncertainties in quadrature to obtain a combined mass constraint of $M_\mathrm{tot} = 1.21\pm0.02\,M_\odot$.

Because we fixed $i_\mathrm{disk}$ and the distance to \uzte\ in these analyses, there are two more sources of statistical uncertainty on $M_\mathrm{tot}$ to be accounted for. As the spatial resolution of the observations degrades, the measurements of $M_\mathrm{tot}$ and $i_\mathrm{disk}$ begin to correlate along the $M_\mathrm{tot} \sin^2 i_\mathrm{disk}$ disk rotation curve degeneracy. To reintroduce the uncertainty associated with $i_\mathrm{disk}$ onto the inference of $M_\mathrm{tot}$, we translate the $\sigma_{i\mathrm{disk}}$=$1.5\degr$ Gaussian inclination uncertainty from \citet{long18} into an $M_\mathrm{tot}$ uncertainty along the $M_\mathrm{tot} \sin^2 i_\mathrm{disk}$ degeneracy, yielding $\sigma_{M\mathrm{tot},i} = 0.05\,M_\odot$. The uncertainty in the distance to the system ($d = 131.2 \pm 1.7\,\mathrm{pc}$) translates linearly into an uncertainty in the total mass ($\sigma_{M\mathrm{tot},d} = 0.01\,M_\odot$). These two uncertainties are added in quadrature to deliver a final mass determination of $M_\mathrm{tot} = 1.21\pm 0.05\,M_\odot$. 
This value agrees well with the \citet{simon00} analysis using \twelve, which obtained $M_\mathrm{tot} = 1.22 \pm 0.07 M_\odot$ (scaled to $d=131.2\,$pc).
Throughout all of the choices of temperature power law (floating, and $q=0.50$, and $q=0.75$), the dynamical mass remained the same for each transition within statistical uncertainties.

\begin{deluxetable}{lcc}
\tablecaption{Inferred Disk Model Parameters\label{table:components}}
\tablehead{\colhead{Parameter} & \thirteen & \eighteen}
\startdata
$M_\mathrm{tot}\quad [M_\odot]$ & $1.19 \pm 0.01$ & $1.23 \pm 0.01$ \\
$r_c$ [au] & $28\pm2$ & $24\pm2$ \\
$T_{10}$ [K] & $48\pm1$ & $28 \pm1$ \\
$q$ & 0.5\tablenotemark{f} & 0.5\tablenotemark{f} \\
$\log_{10} M_\mathrm{disk} \quad \log_{10} [M_\odot]$ & $-5.77\pm0.08$ & $-5.3\pm0.1$ \\
$\xi\,[\kms]$ & $0.23\pm0.02$ & $0.19\pm0.02$ \\
$i_\mathrm{disk} \quad$ [deg] & $56.15$\tablenotemark{f} & $56.15$\tablenotemark{f} \\
$\Omega_\mathrm{disk}$ [deg] & $269.9\pm0.4$ & $269.3\pm0.4$ \\
$v_r$\,$[\kms]$ & $5.74\pm0.01$\tablenotemark{a} & $5.71\pm0.01$\tablenotemark{a} \\
$\delta_\alpha\,['']$  & $0.79\pm0.004$ & $0.78\pm0.004$ \\
$\delta_\delta\,['']$ & $-0.21\pm0.004$ & $-0.21\pm0.004$ \\
\enddata
\tablenotetext{f}{Parameter is fixed.}
\tablenotetext{a}{LSRK reference frame.}
\tablecomments{The 1D marginal posteriors are well-described by a Gaussian, so we report symmetric error bars here (statistical uncertainties only). These parameters were inferred using a distance of $d = 131.2\,$pc.}
\end{deluxetable}

\uzte\ was first discovered as a single-lined spectroscopic binary by \citet{mathieu96}. Using high resolution infrared spectroscopy, \citet{prato02} revealed \uzte\ as a double-lined spectroscopic binary (SB2). \citet{jensen07} further refined the orbital parameters using additional RV observations, including those acquired by \citet{martin05}. Without needing to know the distance to the system, double-lined radial velocity solutions yield the mass ratio of the stars ($q = M_B/M_A$) as well as the quantity $M_\mathrm{tot} \sin^3 i_\star$. Since the protoplanetary-disk based technique independently measures $M_\mathrm{tot}$, we can combine these two results to solve directly for $i_\star$. For \uzte, \citet{jensen07} find $M_\mathrm{tot} \sin^3 i_\star = 0.69 \pm 0.13\,M_\odot$, and so we infer the binary inclination relative to the sky-plane to be $i_\star = 56.1 \pm 5.7\degr$ (technically, there is also a degenerate solution with $i_\star = 123.9 \pm 5.7\degr$). Using \citet{jensen07}'s mass ratio of $q = 0.30\pm0.03$, the individual stellar masses are $M_\mathrm{Ea} = 0.93 \pm 0.04 \,M_\odot$ and $M_\mathrm{Eb} = 0.28 \pm 0.02 \,M_\odot$.


As we noted in the Introduction, calculating the mutual inclination between the binary orbit and the circumbinary disk requires knowledge of $\Omega_\star$. Including \uzte, there are now 4 systems with $i_\mathrm{disk} \simeq i_\star$ but for which our ignorance of $\Omega_\star$ prevents a direct calculation of mutual inclinations. Nevertheless we can make statistical statements, which as we show in the next section are constraining.

\section{The mutual inclinations of circumbinary disks} \label{sec:sample}
In this section, we first compile all circumbinary protoplanetary and debris disks in the literature. Then, we estimate the mutual inclination $\imut$ of these systems. For some systems, $\imut$ can be calculated directly via Equation~\ref{eqn:fekel}, whereas other systems require an indirect approach. For the CB disks around SB2s in particular, we show that a na\"ive estimate of mutual inclination is biased, so we implement a hierarchical Bayesian model to infer the mutual inclination distribution of this subsample. Finally, we examine how the mutual inclination distribution changes with binary orbital period across the full sample.

\subsection{The circumbinary disk sample}
The majority of protoplanetary CB disks were identified in two ways. First, there are those that originated from radial velocity (RV) surveys for spectroscopic binaries in star forming regions \citep[e.g.,][]{mathieu94,melo03,guenther07}. These RV searches were primarily sensitive to binaries with orbits shorter than 1 year (semi-major axes smaller than 1\,au). Some fraction of the binary stars discovered by these surveys were found to have spectroscopic accretion signatures and infrared and mm-excesses above the stellar photosphere. The disks around these binaries were targeted with sub-mm interferometers \citep[e.g.,][]{simon00,rosenfeld12b,czekala15a} or high-contrast imagers \citep{ginski18}.

Second, there are the protoplanetary CB disks that originated from high contrast adaptive optics and/or non-redundant masking (NRM) searches for binary stars and planetary-mass companions in star forming regions \citep[e.g.,][]{ireland08,ruiz-rodriguez16}. Current instrumentation has enabled observations that probe binary orbits with separations as small as 10\,au. These techniques, in addition to infrared interferometry \citep{schaefer18}, are also used to monitor the orbital motions of binary stars and companions. In many cases, spatially resolved sub-mm observations of the discovered binary sources were independently acquired by surveys of these same star forming regions \citep{cox17}.
 
A third, smaller, sample of protoplanetary CB disks are those identified indirectly through light curve analysis. Long-term photometric monitoring campaigns of T~Tauri stars discovered some sources with periodic, evolving dips \citep[e.g. KH~15D;][]{herbst02,winn03,johnson04}. Modeling of these systems indicated that an optically thick, inclined circumbinary disk was responsible for screening the stars \citep{chiang04,plavchan08}. There is also one EB whose variable light curve shows evidence for a circumbinary disk \citep{gillen14,gillen17}. The remainder of the protoplanetary CB disks were identified by miscellaneous means, often serendipitously from detailed studies of individual targets. 

The sample of circumbinary debris disks was acquired in much the same way as the protoplanetary disk sample, with disk discoveries primarily coming from surveys by the \emph{Herschel} satellite at far-IR and sub-mm wavelengths \citep{matthews10} and binarity follow-up observations with adaptive optics. There are a few nearby resolved CB debris disks \citep[e.g.,][]{kennedy12a,kennedy12b,kennedy15}; however, the modest spatial resolution of \emph{Herschel} means that most debris disks known to be circumbinary are spatially unresolved \citep{rodriguez15}.

We separate the full collection of CB disk systems into three categories based upon the fidelity of their measured parameters and their ability to inform our study of the distribution of mutual inclinations. The first group contains the circumbinary systems that have precise measurements of the disk and stellar orientations (Table~\ref{tab:disks-known}) and includes six protoplanetary disks and four debris disks. The second group contains protoplanetary systems which may have partial stellar orbits and/or ambiguities in their disk orientations, which leads to moderate uncertainties in mutual inclinations (Table~\ref{tab:disks-quasi}). The third group contains those protoplanetary systems which are known to be circumbinary, but do not have complete orbit information and/or a spatially resolved disk observation, and so no mutual inclination can be calculated (Table~\ref{tab:disks-unknown}). We include these systems here in the hope that Table~\ref{tab:disks-unknown} may serve as a central repository to motivate future CB follow-up observations. \citet{rodriguez15} provides an additional 30 spatially unresolved debris disks from the DEBRIS survey \citep{matthews10} that are likely to be circumbinary.

\begin{deluxetable*}{lccccccccccccl}
\tablecaption{CB disks around binary stars with precisely measured orbital parameters  \label{tab:disks-known}}
\tablehead{\colhead{Name} & \colhead{$P$} & \colhead{$M_1$} & \colhead{$M_2$} & \colhead{$q$} & \colhead{$a$} & \colhead{$e$} & \colhead{$i_\mathrm{disk}$} & \colhead{$\Omega_\mathrm{disk}$} & \colhead{$i_\star$} & \colhead{$\Omega_\star$} & \colhead{$\imut$} & \colhead{age} & \colhead{References}\\ \colhead{ } & \colhead{days} & \colhead{$M_\odot$} & \colhead{$M_\odot$} & \colhead{ } & \colhead{au} & \colhead{ } & \colhead{$\degr$} & \colhead{$\degr$} & \colhead{$\degr$} & \colhead{$\degr$} & \colhead{$\degr$} & \colhead{Myr} & \colhead{ }}
\startdata
V4046 Sgr\tablenotemark{q} & 2.4 & 0.9 & 0.85 & 0.94 & 0.04 & 0.00 & 33.5$\pm$1.4 & 256$\pm$1.0 & 33.4$\pm$0.6 & \nodata & $<$2.3\tablenotemark{B} & 13 & 1,2,3,4 \\
CoRoT 2239 & 3.9 & 0.67 & 0.495 & 0.74 & 0.05 & 0.00 & 81$\pm$5 & \nodata & 85.09$\pm$0.15 & \nodata & $<$5 & 3 & 5,6,7 \\
AK Sco & 13.6 & 1.25 & 1.25 & 1.00 & 0.16 & 0.47 & 109.4$\pm$0.5 & 51.1$\pm$0.3 & 108.8$\pm$2.4 & 48$\pm$3 & $<$2.7\tablenotemark{B} & 18 & 8,9,10 \\
DQ Tau & 15.8 & 0.63 & 0.59 & 0.94 & 0.15 & 0.57 & 160$\pm$3 & 4.2$\pm$0.5 & 158.2$\pm$2.8 & \nodata & $<$2.7\tablenotemark{B} & 3 & 11 \\
UZ Tau E\tablenotemark{q} & 19.1 & 1.02 & 0.29 & 0.29 & 0.15 & 0.33 & 56.15$\pm$1.5 & 269.6$\pm$0.5 & 56.1$\pm$5.7 & \nodata & $<$2.7\tablenotemark{B} & 3 & 12,13,14 \\
HD 98800 B\tablenotemark{q} & 315.0 & 0.7 & 0.6 & 0.86 & 1.05 & 0.78 & 154$\pm$1 & 197$\pm$2 & 67$\pm$3 & 157.6$\pm$2.4 & 92$\pm$3 & 10 & 15,16,17,18 \\
\hline
HD 131511 & 11.5 & 0.79 & 0.45 & 0.57 & 0.19 & 0.51 & 90$\pm$10 & 245$\pm$5 & 93.4$\pm$4.2 & 248$\pm$3.6 & $<$15 & $10^3$ & 19 \\
$\alpha$ CrB & 17.4 & 2.58 & 0.92 & 0.36 & 0.20 & 0.37 & 90$\pm$10 & 345$\pm$20 & 88.2$\pm$0.1 & 330$\pm$20 & $<$35 & 730 & 20 \\
$\beta$ Tri & 31.4 & 3.5 & 1.4 & 0.40 & 0.31 & 0.43 & 130$\pm$10 & 247$\pm$10 & 130.0$\pm$0.5 & 245.2$\pm$0.67 & $<$14 & 350 & 20 \\
99 Her & 56.3 yr & 0.94 & 0.46 & 0.49 & 16.58 & 0.77 & 45$\pm$5 & 72$\pm$10 & 39$\pm$2 & 41$\pm$2 & 80$\pm$6 & $6\times10^3$ & 21
\enddata
\tablenotetext{B}{Mutual inclination inferred via hierarchical Bayesian model.}
\tablenotetext{t}{Hosts a single companion orbiting beyond the circumbinary disk.}
\tablenotetext{q}{Hosts a binary companion orbiting beyond the circumbinary disk.}
\tablecomments{1) \citet{stempels04}, 2) \citet{rosenfeld12b}, 3) \citet{kastner18a}, 4) \citet{kastner18b}, 5) \citet{gillen14}, 6) \citet{terquem15}, 7) \citet{gillen17}, 8) \citet{alencar03}, 9) \citet{anthonioz15}, 10) \citet{czekala15a}, 11) \citet{czekala16}, 12) \citet{simon00}, 13) \citet{prato02}, 14) \citet{jensen07}, 15) \citet{andrews10a}, 16) \citet{boden05}, 17) \citet{ribas18}, 18) \citet{kennedy19}, 19) \citet{kennedy15}, 20) \citet{kennedy12a}, 21) \citet{kennedy12b}. Mutual inclination $\imut$ upper limits enclose 68\% of the posterior probability distribution. Protoplanetary disks are above the horizontal rule and debris disks are below it.}
\end{deluxetable*}
\begin{deluxetable*}{lccccccccccccl}
\tablecaption{CB protoplanetary disks around stars with moderate orientation uncertainties \label{tab:disks-quasi}}
\tablehead{\colhead{Name} & \colhead{$P$} & \colhead{$M_1$} & \colhead{$M_2$} & \colhead{$q$} & \colhead{$a$} & \colhead{$e$} & \colhead{$i_\mathrm{disk}$} & \colhead{$\Omega_\mathrm{disk}$} & \colhead{$i_\star$} & \colhead{$\Omega_\star$} & \colhead{$\imut$} & \colhead{age} & \colhead{References}\\ \colhead{ } & \colhead{ } & \colhead{$M_\odot$} & \colhead{$M_\odot$} & \colhead{ } & \colhead{au} & \colhead{ } & \colhead{$\degr$} & \colhead{$\degr$} & \colhead{$\degr$} & \colhead{$\degr$} & \colhead{$\degr$} & \colhead{Myr} & \colhead{ }}
\startdata
TWA 3A\tablenotemark{t} & 34.9 d & \nodata & \nodata & 0.84 & 0.127 & 0.628 & 43$\pm$10 & 110$\pm$15 & 47$\pm$15 & 108$\pm$15 & $<$25 & 10 & 1,2,* \\
GW Ori A-B\tablenotemark{c} & 241.5 d & 2.8 & 1.68 & 0.60 & 1.25 & 0.13 & 137$\pm$2 & 1$\pm$1 & 157$\pm$1 & 263$\pm$13 & 50$\pm$5 & 1 & 3 \\
HD 200775 & 3.92 yr & 5.37 & 4.4 & 0.82 & 5 & 0.3 & 55$\pm$1 & 180$\pm$8\tablenotemark{a} & 66$\pm$7 & 172$\pm$6 & $<$20 & 0.1 & 4,5,6 \\
GW Ori AB-C\tablenotemark{c} & 11.0 yr & 4.48 & 1.15 & 0.26 & 9.2 & 0.13 & 137$\pm$2 & 1$\pm$1 & 150$\pm$7 & 282$\pm$9 & 45$\pm$5 & 1 & 3 \\
R CrA & 30 yr & 2 & 0.5 & 0.25 & 29 & 0.4 & 35$\pm$10\tablenotemark{a} & 180$\pm$10\tablenotemark{a} & 70$\pm$15 & \nodata & $>$10 & 1 & 7,8 \\
HD 142527 & 50 yr & 2.1 & 0.11 & 0.05 & $<$50 & 0.5 & 153$\pm$1 & 160.9$\pm$1 & 125$\pm$5 & 130$\pm$10\tablenotemark{h} & 35$\pm$5\tablenotemark{h} & 1 & 9,10,11,12,13 \\
SR 24N\tablenotemark{t} & 111 yr & \nodata & \nodata &  & 25 & 0.64 & 121$\pm$7 & 297$\pm$5 & 132$\pm$6 & 72$\pm$4\tablenotemark{s} & 37$\pm$15\tablenotemark{s} & 1 & 14,15,16 \\
GG Tau Ab1-Ab2\tablenotemark{c} & 14 yr & 0.38 & 0.3 & 0.80 & 4.5 & \nodata & 35$\pm$1 & 277$\pm$0.2 & \nodata & \nodata & \nodata & 3 & 17,18,19,20,21 \\
GG Tau Aa - Ab\tablenotemark{c} & $>$400 yr & 0.6 & 0.68 & 1.13 & 35 & 0.5 & 143$\pm$1 & 277$\pm$0.2 & 132.5$\pm$2 & 133$\pm10$\tablenotemark{g} & 25$\pm$5\tablenotemark{g} & 3 & 17,18,19,20,21 \\
IRS 43 & 450 yr & 1 & 1 & 1.00 & 74 & \nodata & 70$\pm$10 & 90$\pm$5 & $<$30 & \nodata & $>$40 & $<$0.1 & 22
\enddata
\tablenotetext{a}{Ambiguity exists in the orbit orientation ($i < 90\degr$ or $i > 90\degr$) or ($\Omega < 180\degr$ or $\Omega > 180\degr$).}
\tablenotetext{h}{There is a $180\degr$ ambiguity in $\Omega_\star$ due to lack of RV information, so there is a possible solution with $\Omega_\star =310\pm10\degr$ and $\imut = 80\pm10\degr$.}
\tablenotetext{s}{There is a $180\degr$ ambiguity in $\Omega_\star$ due to lack of RV information, so there is a possible solution with $\Omega_\star =252\pm4\degr$ and $\imut = 96\pm20\degr$.}
\tablenotetext{g}{There is a $180\degr$ ambiguity in $\Omega_\star$ due to lack of RV information, so there is a possible solution with $\Omega_\star = 313\pm10\degr$ and $\imut = 80\pm5\degr$.}
\tablenotetext{t}{Hosts a single companion orbiting beyond the circumbinary disk.}
\tablenotetext{c}{System hosts a circumternary disk.}
\tablecomments{1) \citet{andrews10a}, 2) \citet{kellogg17}, 3) \citet{czekala17b}, 4) \citet{monnier06}, 5) \citet{okamoto09}, 6) \citet{benisty13}, 7) \citet{kraus09}, 8) \citet{mesa19}, 9) \citet{biller12}, 10) \citet{lacour16}, 11) \citet{boehler17}, 12) \citet{price18}, 13) \citet{claudi18}, 14) \citet{andrews05}, 15) \citet{fernandez-lopez17}, 16) \citet{schaefer18}, 17) \citet{andrews14}, 18) \citet{difolco14}, 19) \citet{dutrey16}, 20) \citet{tang16}, 21) \citet{cazzoletti17}, 22) \citet{brinch16}, *) Czekala et al. \emph{in prep}. Alternative names: TWA\,3A = Hen\,3-600; HD\,200775 = MWC\,361; SR\,24N = WSB\,41.}
\end{deluxetable*}
\begin{deluxetable*}{lcccccccccccl}
\tablecaption{CB protoplanetary disks with unknown orientations.\label{tab:disks-unknown}}
\tablehead{\colhead{Name} & \colhead{$P$} & \colhead{$M_1$} & \colhead{$M_2$} & \colhead{$q$} & \colhead{$a$} & \colhead{$e$} & \colhead{$i_\mathrm{disk}$} & \colhead{$\Omega_\mathrm{disk}$} & \colhead{$i_\star$} & \colhead{$\Omega_\star$} & \colhead{age} & \colhead{References}\\ \colhead{ } & \colhead{ } & \colhead{$M_\odot$} & \colhead{$M_\odot$} & \colhead{ } & \colhead{au} & \colhead{ } & \colhead{$\degr$} & \colhead{$\degr$} & \colhead{$\degr$} & \colhead{$\degr$} & \colhead{Myr} & \colhead{ }}
\startdata
HD 104237 & 20 d & 2.2 & 1.4 & 0.64 & 0.22 & 0.6 & \nodata & \nodata & $>$90 & 235$\pm$3 & 2 & 1,2 \\
HD 34700A\tablenotemark{t} & 23.5 d & 2 & 2 & 0.99 & 0.69 & 0.25 & 42 & 86 & \nodata & \nodata & 5 & 3,4 \\
ROXs 42Ca\tablenotemark{t} & 36 d & \nodata & \nodata & 0.91 & \nodata & 0.48 & \nodata & 116$\pm$4\tablenotemark{a} & \nodata & \nodata & 2 & 5,6,7,8,9,10 \\
CD-22 11432 & 36 d & 1 & 1 & 1.00 & \nodata & \nodata & \nodata & \nodata & \nodata & \nodata & 8 & 11,* \\
GV Tau S\tablenotemark{t} & 38 d & 0.5 & 0.13 & 0.26 & \nodata & \nodata & 55 & 160 & \nodata & \nodata & 0.4 & 12,13,14,15 \\
KH 15D & 48.4 d & 0.715 & 0.74 & 1.03 & 0.29 & 0.57 & 84$\pm$2\tablenotemark{j} & 107$\pm$1\tablenotemark{j} & 92.5$\pm$2.5 & \nodata & 4 & 16,17,18,19,20,21,22 \\
HD 106906 & 49.2 d & 1.37 & 1.34 & 0.98 & \nodata & 0.67 & 85 & 104\tablenotemark{a} & 88 & \nodata & 15 & 23,24 \\
YLW 16A\tablenotemark{t} & 92.6 d & \nodata & \nodata &  & \nodata & \nodata & \nodata & \nodata & \nodata & \nodata & 1 & 25 \\
AS 205S\tablenotemark{t} & \nodata & 0.74 & 0.54 & 0.73 & \nodata & \nodata & 66$\pm$2 & 110$\pm$2 & \nodata & \nodata & 0.5 & 26,11,27 \\
WL 4\tablenotemark{t} & 130.87 d & \nodata & \nodata &  & \nodata & \nodata & \nodata & \nodata & \nodata & \nodata & 1 & 28,29 \\
WSB 74 & $<$150 d & 0.86 & 0.817 & 0.95 & $<$0.6 & \nodata & \nodata & \nodata & \nodata & \nodata & 0.1 & 30,31,9 \\
WSB 40 & \nodata & 0.96 & 0.75 & 0.78 & 2.3 & \nodata & \nodata & 167$\pm$32\tablenotemark{a} & \nodata & \nodata & 1.5 & 31,9 \\
V935 Sco & \nodata & 1.11 & 0.75 & 0.68 & 2.6 & \nodata & \nodata & 81$\pm$5\tablenotemark{a} &  &  & 2.1 & 31,9 \\
V835 Oph Aa-Ab\tablenotemark{t} & \nodata & \nodata & \nodata &  & 3.2 & \nodata & \nodata & 90$\pm$27\tablenotemark{a} & \nodata & \nodata & 2 & 28,9,10 \\
ROph 36 & \nodata & 0.73 & 0.61 & 0.84 & 3.3 & \nodata & \nodata & 77$\pm$15\tablenotemark{a} & \nodata & \nodata & 2 & 32,33,31,9 \\
CS Cha & 7 yr & \nodata & \nodata &  & 4.0 & \nodata & 24$\pm$3\tablenotemark{a} & 75$\pm$2\tablenotemark{a} & \nodata & \nodata & 2 & 34,35,36,37 \\
V892 Tau & 14 yr & 2.25 & 2.25 & 1.00 & 7 & 0.12 & $>$60 & 49$\pm$1\tablenotemark{a} & 60$\pm$4 & 28$\pm$5\tablenotemark{a} & $<$3 & 38,39 \\
MHO 2AB & \nodata & 0.33 & 0.11 & 0.34 & 7.3 & \nodata & \nodata & \nodata & \nodata & \nodata & 3 & 40,41 \\
CoKu Tau/4 & \nodata & 0.5 & 0.5 & 1.00 & 8 & \nodata & \nodata & \nodata & \nodata & \nodata & 4 & 42,43,44 \\
IC 348 LRL 31 & \nodata & 1.62 & 0.2 & 0.12 & 8.4 & \nodata & \nodata & \nodata & \nodata & \nodata & 8 & 31,45 \\
VLA 1623A\tablenotemark{t} & \nodata & \nodata & \nodata &  & 30 & \nodata & \nodata & 20\tablenotemark{a} & \nodata & \nodata & $<$0.1 & 46 \\
L1551 IRS 5 & \nodata & 0.65 & \nodata &  & 70 & \nodata & 62 & 167 & \nodata & \nodata & 0.1 & 47 \\
L1448 IRS3B\tablenotemark{t} & \nodata & \nodata & \nodata &  & 72 & \nodata & 45 & 30 & \nodata & \nodata & $<$1 & 48 \\
HH 250 & \nodata & \nodata & \nodata &  & 120 & \nodata & \nodata & \nodata & \nodata & \nodata & $<$1 & 49 \\
IRAS17216-3801 & \nodata & \nodata & \nodata &  & 170 & \nodata & \nodata & \nodata & \nodata & \nodata & $<$1 & 50 \\
UY Tau & 1640 yr & \nodata & \nodata &  & 190 & \nodata & 50$\pm$10 & 42$\pm$3 & \nodata & \nodata & 3 & 51,52,53
\enddata
\tablenotetext{j}{Inclination and position angle determined from the outflow jet assuming it is orthogonal to the CB disk. There is a 90 degree ambiguity in $i_\mathrm{disk}$ and a 180 ambiguity in $\Omega_\mathrm{disk}$.}
\tablecomments{Table superscripts are the same as in Table~\ref{tab:disks-quasi}. References: 1) \citet{bohm04}, 2) \citet{garcia13}, 3) \citet{torres04}, 4) \citet{monnier19}, 5) \citet{mathieu89}, 6) \citet{ghez93}, 7) \citet{lee94}, 8) \citet{barsony03}, 9) \citet{cox17}, 10) \citet{schaefer18}, 11) \citet{barenfeld16}, 12) \citet{menard93}, 13) \citet{doppmann08}, 14) \citet{guilloteau11}, 15) \citet{sheehan14}, 16) \citet{hamilton01}, 17) \citet{hamilton03}, 18) \citet{johnson04}, 19) \citet{chiang04}, 20) \citet{hamilton05}, 21) \citet{winn06}, 22) \citet{aronow18}, 23) \citet{kalas15}, 24) \citet{derosa19}, 25) \citet{plavchan13}, 26) \citet{eisner05}, 27) \citet{kurtovic18}, 28) \citet{ratzka05}, 29) \citet{plavchan08}, 30) \citet{kohn16}, 31) \citet{ruiz-rodriguez16}, 32) \citet{cieza10}, 33) \citet{orellana12}, 34) \citet{guenther07}, 35) \citet{espaillat11}, 36) \citet{dunham16}, 37) \citet{ginski18}, 38) \citet{smith05}, 39) \citet{monnier08}, 40) \citet{kraus11}, 41) \citet{harris12}, 42) \citet{dalessio05}, 43) \citet{ireland08}, 44) \citet{nagel10}, 45) \citet{ruiz-rodriguez18}, 46) \citet{harris18}, 47) \citet{takakuwa17}, 48) \citet{tobin16}, 49) \citet{comeron18}, 50) \citet{kraus17}, 51) \citet{close98}, 52) \citet{hioki07}, 53) \citet{tang14}, *) G. Torres, \emph{private communication}. Alternative names: HD\,104237=DX\,Cha; ROXs\,42C=ROph\,26, NTTS\,162814-2427; CD-22\,11432=2MASS\,J16141107-2305362, GV\,Tau\,S=Haro\,6-10\,N; AS\,205S=AS\,205B, 2MASS\,J16113134-1838259; YLW 16A=2MASS\,J16272802-2439335; V380\,Ori\,A=HH\,222; WL\,4=2MASS\,J16271848-2429059, ISO-Oph\,128; WSB\,74=ROph 32, J16315473-2503238; WSB\,40=ROph\,12; V935\,Sco=ROph\,2, WSB\,12; V835\,Oph\,A=SR\,13, ROph\,23; ROph\,36=2MASS J16335560-2442049\,AB; HH\,250=IRAS\,19190+1048; V892\,Tau=Elias\,1.}
\end{deluxetable*}

Many of the circumbinary disks reside in hierarchical triple or quadruple systems, where the companions orbit exterior to the CB disk. We denote these with superscripts in the tables. The frequency of tertiary companions to close spectroscopic binaries is known to be a strong function of binary period: 96\% of binaries with periods $P < 3$ days have an outer companion, whereas only 34\% of binaries with $12 < P < 30$ days have a tertiary \citep{tokovinin06}; the latter percentage is consistent with the tertiary rate of our sample (Table~\ref{tab:disks-known}). There are also two circumternary systems, GW~Ori and GG~Tau~A, which have a single disk encircling all three stars.

\subsection{Direct and indirect measurements of mutual inclination}
\label{subsec:standard-bayes}
If there are complete measurements of the disk and binary orientations $i_\mathrm{disk}$, $\Omega_\mathrm{disk}$, $i_\star$, $\Omega_\star$, then the mutual inclination can be calculated directly via Equation~\ref{eqn:fekel}. The debris disks in Table~\ref{tab:disks-known} are sufficiently nearby that complete astrometric observations of the stellar orbit exist (providing $\Omega_\star$) and the mutual inclination can be determined directly and unambiguously. There are only two protoplanetary disks with sufficiently complete astrometric observations and spatially resolved disk observations for which a similarly precise calculation is possible: \ak\ and HD~98800B. One other protoplanetary systems, CoRoT~223992193 (hereafter CoRoT~2239) does not have spatially resolved disk observations or astrometric orbits; however, its mutual inclination has been inferred from photodynamical modeling of its light curve and RVs. 

There are several spatially resolved CB protoplanetary disks with astrometric observations in Table~\ref{tab:disks-quasi}, but insufficient orbital phase coverage leaves moderate (often correlated) uncertainties in $i_\star$ and $\Omega_\star$. Because the transformation into mutual inclination is non-linear, these systems have substantial mutual inclination uncertainties. Even so, near-coplanar mutual inclinations (i.e., $\imut \lesssim 10\degr$) are ruled out at high significance for many of these systems: GW~Ori, HD~142572, SR~24N, and IRS~43.

The three remaining CB systems, \vsgr, \dq, and \uzte, contain double-lined spectroscopic binaries (SB2s) and have precise measurements of $i_\mathrm{disk}$, $\Omega_\mathrm{disk}$, and $i_\star$, but not $\Omega_\star$. Because we measure $i_\mathrm{disk} \simeq i_\star$ for each of the three disks in this subsample, we might na\"ively conclude that their $\imut$ values are small. In the remainder of this subsection (\S\ref{subsec:standard-bayes}), we use a standard Bayesian analysis to demonstrate why this supposition is incorrect when each system is considered individually. At the same time, our intuition tells us that it would be strange to live in a universe where $\imut$ is broadly distributed and yet find $i_\mathrm{disk} \simeq i_\star$ for all three systems. In the upcoming subsection (\S\ref{subsec:hbm}), we introduce a hierarchical Bayesian model that considers all of the CB disks in the subsample to infer a $\imut$ distribution that is indeed narrow.

\paragraph{Biased measurement of $\imut$ for an individual CB disk around an SB2}
Let $\boldsymbol{\kappa} = \{i_\mathrm{disk}, \Omega_\mathrm{disk}, i_\star \}$ and let $\boldsymbol{D}$ denote the measurements of these parameters. Let $\boldsymbol{\lambda} = \{\Omega_\star\}$ represent the parameters we do not measure directly, so that the full vector of parameters is given by $\boldsymbol{\mu} = \{\boldsymbol{\kappa}, \boldsymbol{\lambda}\}$. To obtain the posterior distribution of $\imut$ given $\boldsymbol{D}$, we first derive a posterior distribution of $\boldsymbol{\mu}$. Then, we use Equation~\ref{eqn:fekel} to transform samples from this posterior into samples of the posterior distribution $\theta(\boldsymbol{\mu})$. The likelihood is a multivariate normal 
\begin{equation}
{\cal L} = {\cal N}( \boldsymbol{D} | \boldsymbol{\kappa}, \boldsymbol{\Sigma})
\end{equation}
where the measurement uncertainties on $\boldsymbol{D}$ are described by the covariance matrix $\boldsymbol{\Sigma}$. 

\begin{figure}[tb]
\begin{center}
  \includegraphics{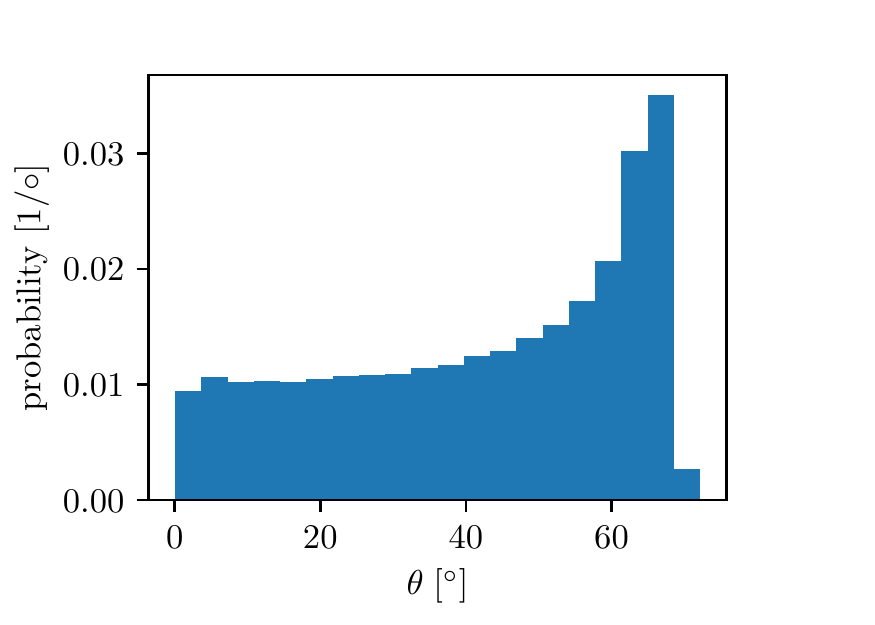}
  \figcaption{Na\"ive posterior distribution of $\imut$ for \vsgr, assuming a isotropic prior distribution for the binary. Similar posterior distributions exist for the other CB disks around SB2s in Table~\ref{tab:disks-known} when treated on an individual basis.
  \label{fig:imut}}
  \end{center}
\end{figure}

For an individual disk, we assume a prior distribution on $\boldsymbol{\mu}$ that specifies the disk and the binary orbit as isotropically oriented in 3D space, i.e., the unit angular momentum vector of each orbit has uniform probability of pointing anywhere on the unit sphere. This is 
\begin{equation}
p(\boldsymbol{\mu}) = \frac{\sin i_\mathrm{disk} \sin i_\star}{4},
\end{equation}
where the prior densities of $\Omega_\mathrm{disk}$ and $\Omega_\star$ are uniform $\in [0, 2\pi]$. We combine the likelihood function with the prior distribution to yield an (unnormalized) posterior probability distribution 
\begin{equation}
p(\boldsymbol{\mu} | \boldsymbol{D}) \propto 
p(\boldsymbol{D} | \boldsymbol{\kappa})\, p(\boldsymbol{\mu}).
\end{equation}
This posterior distribution is sampled using an MCMC algorithm to generate samples of $\boldsymbol{\mu}$, which are then transformed into $\imut$. For the parameters of \vsgr\ listed in Table~\ref{tab:disks-known}, we obtain the posterior in Figure~\ref{fig:imut}.

Even though \vsgr\ has nearly identical disk and binary inclinations (relative to the sky-plane), the posterior distribution has a wide range of permissible $\imut$, telling us that the expected value of $\imut$ is actually much larger than what we would naively assume from $i_\mathrm{disk} \simeq i_\star$. Without a measurement of $\Omega_\star$, the measurement of $i_\star$ simply constrains the binary vector to point within a thin annulus on the surface of the unit sphere, the width of which is set by the measurement uncertainty (see Figure~\ref{fig:observer-frame} for a schematic of such a setup).

While this posterior distribution makes geometrical 
sense when considering an individual CB system in isolation, when applied to a sample of CB systems that all exhibit $i_\mathrm{disk} \simeq i_\star$, the statement that $\imut$ is broadly distributed runs counter to intuition. Even though the expected value of the mutual inclination for any individual system is large (e.g., Figure~\ref{fig:imut}), we would suspect that $\imut$ is narrowly distributed near zero; otherwise we would observe many systems with $i_\mathrm{disk} \not\simeq i_\star$. The following subsection confirms this suspicion in a statistically rigorous way.

\subsection{Inferring the mutual inclinations of CB disks around SB2s with a hierarchical Bayesian model}
\label{subsec:hbm}
Using hierarchical Bayesian analysis, we can explicitly build the mutual inclination distribution (or a parameterization thereof) into our model by considering all disks in the subsample together. Hierarchical Bayesian approaches are useful when there is a natural multilevel structure to a dataset (see \citet{loredo13} and \citet{sharma17} for general introductions in the astrophysical context). Notable applications of hierarchical analysis within the stellar and exoplanetary subfield include inferring the eccentricity distribution of exoplanetary orbits \citep{hogg10b}, the composition distribution of sub-Neptune planets \citep{wolfgang15}, and trends in the stellar obliquity distribution \citep{munoz18}.

To implement this hierarchical model we
\begin{enumerate}
    \item define a flexible parameterization of the mutual inclination distribution $p(\imut)$
    \item simplify the geometrical relationships between $i_\mathrm{disk}$, $i_\star$, and $\theta$ by rotating to a frame centered on the disk
    \item build the full posterior distribution for a sample of $N$ disks
    \item demonstrate the flexibility of the model by using it to correctly infer the mutual inclination distributions of two very different mock samples of disks
    \item apply the framework to the real subsample of CB disks around SB2s
\end{enumerate}

\begin{figure}[tb]
\begin{center}
  \includegraphics{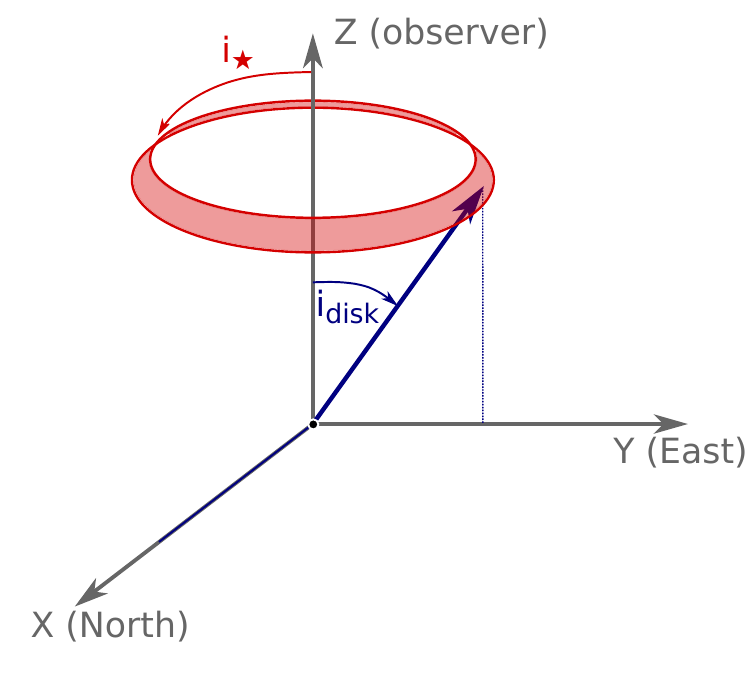}
  \figcaption{If only the inclination of the binary relative to the sky plane $i_\star$ is known (and there is no constraint on $\Omega_\star$), then the orbit normal of the binary can point anywhere in the red annulus with equal probability. This gives rise to mutual inclination posterior probability distributions like those seen in Figure~\ref{fig:imut}.
  \label{fig:observer-frame}}
  \end{center}
\end{figure}

\paragraph{Parameterizing the mutual inclination distribution} 
The mutual inclination distribution (at the top level of the hierarchy) can be thought of as a prior $p(\imut)$ on the mutual inclination value $\imut_j$ for an individual disk $j$ (at the bottom level of the hierarchy). For example, if $p(\imut)$ favored low mutual inclinations, then for a given disk with $i_\mathrm{disk} \simeq i_\star$, the posterior distribution $p(\imut_j |\,\boldsymbol{D}_j)$ would also favor a low mutual inclination (we would also find $\Omega_\mathrm{disk} \simeq \Omega_\star$). However, if $p(\imut)$ favored a broad range of mutual inclinations, then we might end up with a posterior distribution $p(\imut_j |\,\boldsymbol{D}_j)$ more similar to Figure~\ref{fig:imut}. 

To simplify the inference process, we assume a functional form for $p({\theta})$. A necessary quality of a mutual inclination distribution is that it is defined over the range $\imut \in (0,\pi)$ and obeys $\lim_{\theta \to 0} p(\theta) = 0$ and $\lim_{\theta \to \pi} p(\theta) = 0$, since exactly aligned and anti-aligned vectors constitute a set of zero measure. There are many classes of functions that could be used---if the data are sufficiently constraining then the exact choice of functional form will not matter, so long as the function is sufficiently flexible to assume the morphology of the actual mutual inclination distribution. After experimenting with various functional forms, we chose the logit-normal, which provides a wide range of shapes covering the extremes of what we might imagine the mutual inclination distribution to be (i.e., favoring aligned, anti-aligned, or isotropic orientations), while permitting smooth transitions between them. To implement this distribution, we first use the ``logit'' transformation \citep[e.g.,][]{gelman14} to convert $\imut$ from a bounded domain ($0,\pi$) to an intermediate variable $v$ on an unbounded domain ($-\infty, \infty$), 
\begin{equation}
    v = \mathrm{logit}(\imut/\pi) = \ln \left ( \frac{\imut/\pi}{1 - \imut/\pi} \right).
\end{equation}
For each system $j$, we say that $v_j$ is drawn from the distribution 
\begin{equation}
    v_j \sim \, {\cal N}(\mu_v, \tau_v),
\end{equation}
which is a normal distribution with mean $\mu_v$ and precision $\tau_v = 1/\sigma_v^2$, where $\sigma_v$ is the standard deviation of the normal. Together, we call these the hyperparameters of the mutual inclination distribution $\boldsymbol{\alpha} = \{\mu_v, \tau_v\}$. The normal distribution on $v_j$ is equivalent to a prior probability distribution on $\imut_j$ of 
\begin{multline}
    p(\imut_j|\,\boldsymbol{\alpha}) = \frac{1}{\theta (1 - \theta/\pi)}\sqrt{\frac{\tau_v}{2 \pi}} \times \\
    \exp \left (-\frac{\tau_v}{2} \left ( \mathrm{logit}(\imut/\pi) - \mu_v \right)^2 \right)
    \label{eqn:logit-normal}
\end{multline}
We also experimented with different functional forms for $p(\theta |\,\boldsymbol{\alpha})$ including reparameterized Beta functions, and found our results to be unchanged.

\begin{figure}[tb]
\begin{center}
  \includegraphics{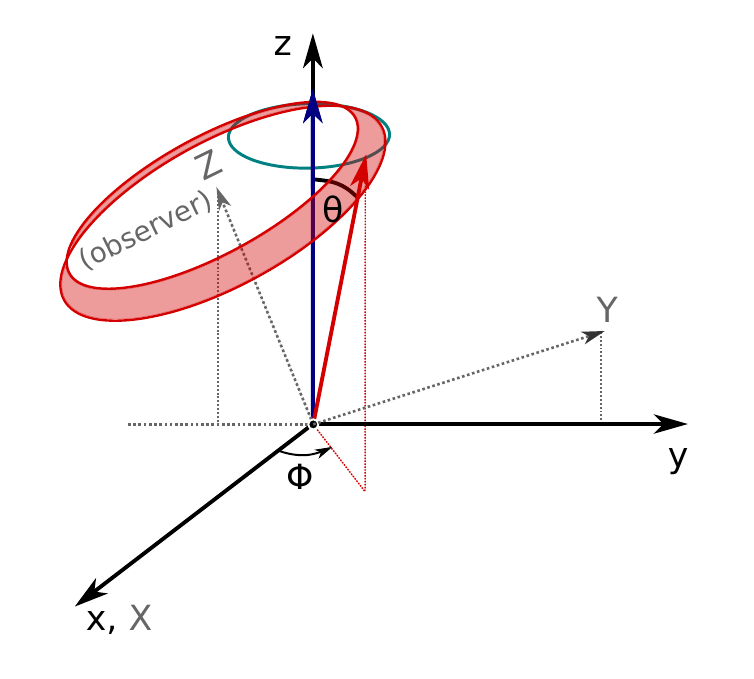}
  \figcaption{The $xyz$ coordinate system in the frame of the circumbinary disk, where the $z$ axis is aligned with the disk unit angular momentum vector. As before, the blue and red vectors are the orbit normals of the disk and binary, respectively. For a given $\imut$, the red annulus specifies the constraint on constant $i_\star$ as in Figure~\ref{fig:observer-frame}. The teal circle represents a constant value of $\imut$.
  \label{fig:disk-frame}}
  \end{center}
\end{figure}

\paragraph{Rotating into the frame of the disk} In Figure~\ref{fig:disk-frame}, we introduce a new $xyz$ coordinate system in the frame of the circumbinary disk, which simplifies the mathematical relationships between $\imut$ and the disk and binary orbit normals. In this frame, the disk angular momentum unit vector is aligned with the $z$ axis, $\imut$ is the polar angle of the binary unit angular momentum vector, and $\phi$ denotes the azimuthal angle of the binary vector. The relationship between the observer frame $XYZ$ (Figures~\ref{fig:coords} and \ref{fig:observer-frame}) and the disk frame $xyz$ (Figure~\ref{fig:disk-frame}) is defined by two rotations about the $Z$ and $x$ axes by the angles $\Omega_\mathrm{disk}$ and $i_\mathrm{disk}$, respectively. Because $\Omega_\star$ is unmeasured, the specific value of $\Omega_\mathrm{disk}$ is irrelevant to the calculation and so we set $\Omega_\mathrm{disk} = 0$ to simplify the required operations (as in Figure~\ref{fig:observer-frame}), which also means $X=x$. Then, all we require is the rotation matrix
\begin{equation}
  \boldsymbol{P}_x(i_\mathrm{disk}) = \left [
  \begin{array}{ccc}
    1 & 0 & 0 \\
    0 & \cos i_\mathrm{disk} & - \sin i_\mathrm{disk} \\
    0 & \sin i_\mathrm{disk} & \cos i_\mathrm{disk} \\
    \end{array}\right]
\end{equation}
which is defined such that its application results in a clockwise rotation of the axes as viewed from $+x$ axis.

For a given $\theta$ and $\phi$, the location of the binary vector in the disk frame (Figure~\ref{fig:disk-frame}) is 
\begin{equation}
\left [ \begin{array}{c}
x\\
y\\
z\\
\end{array} \right]
=
\left [ \begin{array}{c}
\sin \theta \cos \phi \\
\sin \theta \sin \phi \\
\cos \theta \\
\end{array} \right].
\end{equation}
For a specific value of $\theta$, then, the location of the binary vector is constrained to a ring at a constant angular distance from the $z$ axis (teal circle, Figure~\ref{fig:disk-frame}). The most probable locations along the ring (given by $\phi$) are those that coincide with the measurement of $i_\star$, which is represented by the inclined red annulus in Figure~\ref{fig:disk-frame}. To calculate $i_\star$ from ($\theta$, $\phi$) along the ring requires the inverse of the rotation matrix
\begin{equation}
\left [ \begin{array}{c}
X\\
Y\\
Z\\
\end{array} \right]
=
\boldsymbol{P}_x^{-1}(i_\mathrm{disk}) 
\left [ \begin{array}{c}
x\\
y\\
z\\
\end{array} \right]
\end{equation}
which is simply the transpose of $\boldsymbol{P}_x$. Then, $i_\star$ is the angle between the $Z$ axis and the binary orbit normal, so that
\begin{equation}
\cos i_\star = \cos i_\mathrm{disk} \cos \theta - \sin i_\mathrm{disk} \sin \theta \sin \phi. \label{eqn:cosi}
\end{equation}

The main benefit of working in the disk frame is that the prior on the binary orientation is separable,
\begin{equation}
p(\theta_j, \phi_j) = p(\theta_j) p(\phi_j)    
\end{equation}
where $p(\theta)$ is specified by Equation~\ref{eqn:logit-normal} and $p(\phi)$ is uniform $\in [0, 2\pi]$. Equation~\ref{eqn:cosi} defines a relationship between these parameters and $\cos i_\star$, enabling us to write
\begin{multline}
    p(\cos i_\star |\, \cos i_\mathrm{disk}, \theta, \phi) = \\
    \delta \big( \cos i_\star - (\cos i_\mathrm{disk} \cos \theta - \sin i_\mathrm{disk} \sin \theta \sin \phi)\big)
    \label{eqn:delta}
\end{multline}
Technically there are two values of $\phi \in [0, 2\pi]$ that yield the same $\cos i_\star$. Because of the symmetry of the annulus and ring across the $y-z$ plane, we can make Equation~\ref{eqn:delta} a one-to-one relationship by limiting the range to $\phi \in [-\pi/2, \pi/2]$ without loss of generality. 

\paragraph{The full posterior distribution for the CB subsample}
Let $\boldsymbol{\kappa}_j = \{ \cos i_{\mathrm{disk}, j}, \cos i_{\star, j}\}$,  $\boldsymbol{D}_j = \{\cos i_{\mathrm{disk}, j}, \cos i_{\star, j} \}$, $\boldsymbol{\lambda}_j = \{\theta_j, \phi_j \}$, and $\boldsymbol{\mu}_j = \{ \boldsymbol{\kappa}_j, \boldsymbol{\lambda}_j\}$. The posterior probability distribution for the parameters of an individual disk is 
\begin{equation}
    p_j(\boldsymbol{\mu}_j |\, \boldsymbol{D}_j, \boldsymbol{\alpha}) \propto {\cal N}(\boldsymbol{D}_j |\, \boldsymbol{\kappa}_j, \boldsymbol{\Sigma}_j) p(\boldsymbol{\mu}_j | \boldsymbol{\alpha}).
\end{equation}
As before, the likelihood function is evaluated with the observed parameters while the parameters in $\boldsymbol{\lambda}_j$ are constrained by the prior from the mutual inclination distribution and the functional relationship between $\theta, \phi$ and $\cos i_\star$. The prior is
\begin{multline}
    p(\boldsymbol{\mu}_j |\, \boldsymbol{\alpha}) = \\
    p(\cos i_{\star,j} |\, \cos i_{\mathrm{disk},j}, \theta_j, \phi_j) \times \\
     p(\cos i_{\mathrm{disk},j}) \, p(\imut_j |\, \boldsymbol{\alpha}) \, p(\phi_j) 
\end{multline}
where $p(\cos i_{\mathrm{disk},j})$ is a geometrical prior as before. Note that the posterior distribution for an individual disk is conditional on the values within $\boldsymbol{\alpha}$ (which control whether the mutual inclination distribution favors low, isotropic, or high values of $\imut$), reflecting the hierarchical nature of the problem. 

\begin{figure}[tb]
\begin{center}
  \includegraphics{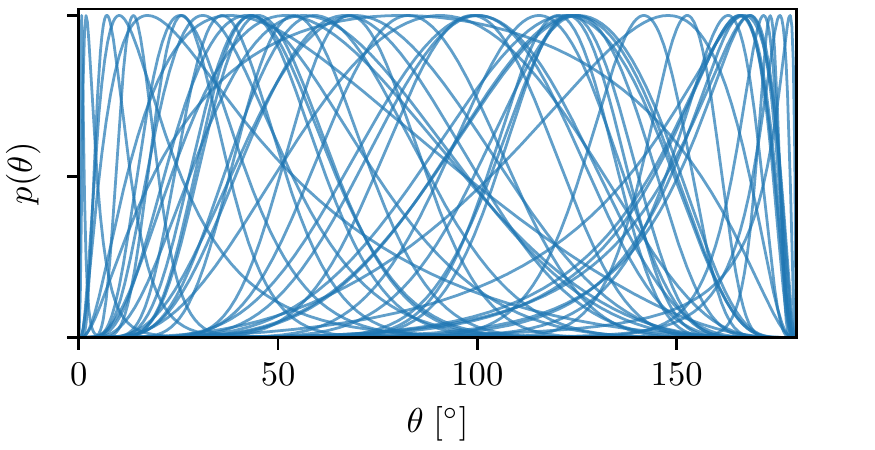}
  \figcaption{The range of mutual inclination distributions $p(\theta|\,\boldsymbol{\alpha})$ randomly drawn from the hyperpriors $p(\boldsymbol{\alpha}|\,\boldsymbol{\beta})$, for a fixed choice of hyperprior parameters $\boldsymbol{\beta}$. To highlight the range of morphologies allowed, each draw has been scaled such that its peak value is one. The range of functional forms demonstrate that the logit-normal distribution can easily mimic distributions that favor mostly aligned, anti-aligned, or isotropic mutual inclinations and easily transition between them.
  \label{fig:prior}}
  \end{center}
\end{figure}

The full posterior distribution is given by 
\begin{multline}
    p \left ( \{\boldsymbol{\mu}_j\}_{j=1}^N, \boldsymbol{\alpha} |\, \{ \boldsymbol{D}\}_{j=1}^N  \right) = \\
    \prod_j^N p_j(\boldsymbol{\mu}_j |\, \boldsymbol{D}_j, \boldsymbol{\alpha}) \times p(\boldsymbol{\alpha} |\, \boldsymbol{\beta})
    \label{eqn:full-posterior}
\end{multline}
where $N$ is the number of systems in the sample. This posterior distribution contains $N \times 3 + 2$ parameters that need to be sampled---the orientation parameters for each system, plus the two hyperparameters of the mutual inclination distribution in $\boldsymbol{\alpha}$. To form a proper posterior probability distribution, the hyperparameters also require their own hyperpriors $p(\boldsymbol{\alpha} |\, \boldsymbol{\beta})$, where $\boldsymbol{\beta}$ are the settings of the hyperprior distribution. These hyperpriors on $\boldsymbol{\alpha}$ are simply chosen such that the range of functional realizations from the logit-normal distribution covers the range of distributions which we hope to infer, without falling victim to pathologies in implementation. From visual inspection of various functional realizations, we decide on a Gaussian prior on $\mu_v$ of $p(\mu_v) \propto {\cal N}(\mu_v |\, \mu_\mu=0, \sigma_\mu=2)$. While the precision parameter $\tau_v$ is already defined to be positive, we find that for very low values ($\tau < 0.5$), the mutual inclination distribution $p(\imut |\, \boldsymbol{\alpha})$ becomes multi-modal with peaks near 0 and $\pi$. To avoid this behavior, we enforce a half-Gaussian prior $p(\tau_v) \propto {\cal N}_{1/2}(\tau_v |\, \mu_\tau = 0.5, \sigma_\tau = 4)$ where $\tau > 0.5$, otherwise $p(\tau_v) = 0$. We refer to this collection of hyperprior parameters as $\boldsymbol{\beta} = \{\mu_\mu, \sigma_\mu, \mu_\tau, \sigma_\tau \}$ and their values remain fixed throughout the entire inference process. To demonstrate the range of possible mutual inclination distributions under this choice of prior, we show random samples from the hyperprior in Figure~\ref{fig:prior}.

\begin{figure}[tb]
\begin{center}
  \includegraphics{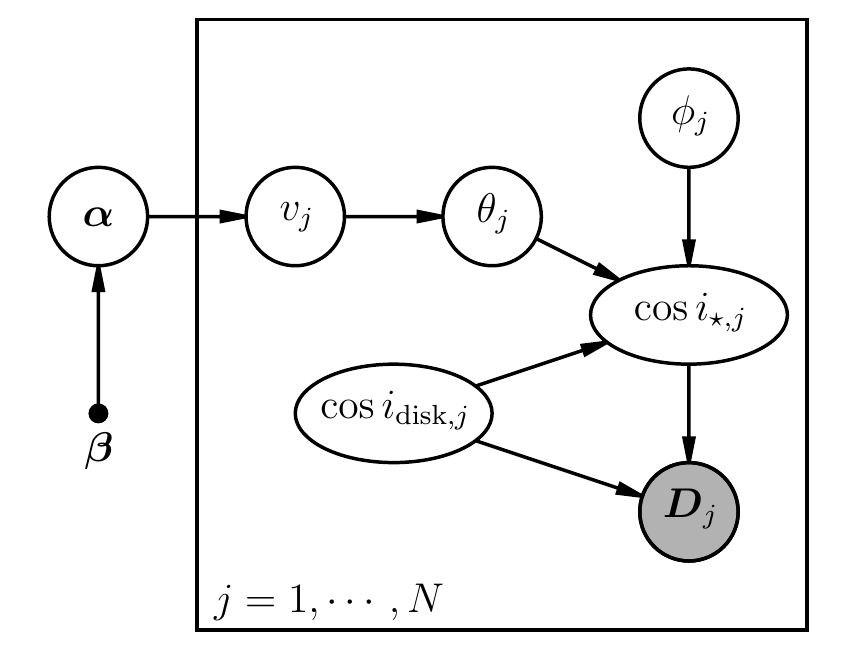}
  \figcaption{The probabilistic graphical model representing the causal relationships between the parameters in our hierarchical Bayesian model. The hyperprior parameters $\boldsymbol{\beta}$ (fixed) describe the range of possible mutual inclination distributions for all disks in the sample (parameterized by $\boldsymbol{\alpha}$, Equation~\ref{eqn:logit-normal}). Within the plate are parameters representing each individual system $j$. The intermediate mutual inclination variable $v_j$ is drawn from the mutual inclination distribution and converted to $\imut_j = \pi\,\mathrm{logit}^{-1} (v_j)$. Then, $\cos i_{\mathrm{disk},j}$ and $\phi_j$ are drawn from their prior distributions and Equation~\ref{eqn:delta} is used to calculate $\cos i_{\star,j}$. Finally, the likelihood for each system is evaluated via the agreement of $\boldsymbol{\kappa}_j = \{\cos i_{\mathrm{disk},j}, \cos i_{\star,j}\}$ with the observed values in $\boldsymbol{D}_j$.
  \label{fig:pgm}}
  \end{center}
\end{figure}

In Figure~\ref{fig:pgm} we show the probabilistic graphical model representing this posterior distribution (Equation~\ref{eqn:full-posterior}). Each node in the graph represents a random variable with a probability distribution attached to it. Arrows between nodes represent causal relationships between parameters. The box or ``plate'' represents the $N$ systems in our sample, and so there are individual nodes for each system for all variables within the plate. 

\paragraph{Testing the model with two mock CB disk samples}
To demonstrate the flexibility of this hierarchical Bayesian model we apply it to two $N$=10 samples of fake circumbinary disk systems with drastically different mutual inclination distributions. The first sample contains disks and binaries with isotropic orientations, i.e., $p_\mathrm{iso}(\imut) \propto \sin(\imut)$. The second sample utilizes the same disk orientations, but the binary orientations have been drawn from a mutual inclination distribution favoring aligned orientations, $ p_\mathrm{low}(\imut) = {\cal N}(\imut |\, \mu_\imut, \sigma_\imut) \sin \imut$, which is a Gaussian with $\mu_\imut = 5\degr$ and $\sigma_\imut = 2\degr$, tapered by a $\sin(\theta)$ profile to satisfy the limit condition. We generate $\theta$ samples from this distribution using rejection sampling \citep[see][Ch 29.3]{mackay03}, draw samples of $\phi$ uniform in the range $[0, 2\pi]$, and convert these $\imut,\phi$ samples to $\cos i_\star$ using Equation~\ref{eqn:delta}. We assume a $1\degr$ uncertainty on $i_\mathrm{disk}$ and $i_\star$ measurements. Inspecting these two samples of fake systems, our intuition from earlier is confirmed by the fact that the majority of disks in the isotropic sample have $i_\mathrm{disk} \not\simeq i_\star$ while all disks in the low mutual inclination sample have $i_\mathrm{disk} \simeq i_\star$.

\begin{figure*}[t]
\begin{center}
  \includegraphics{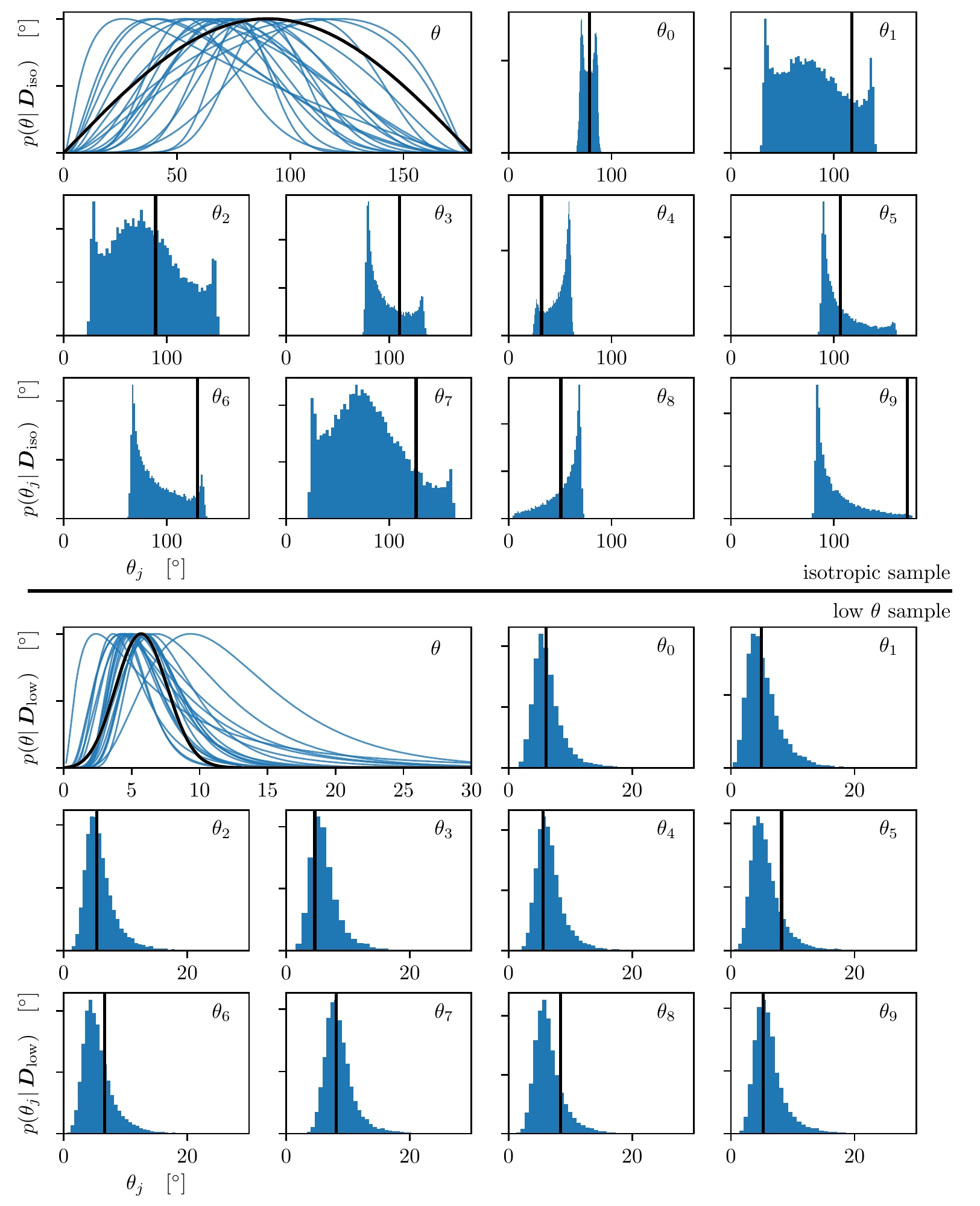}
  \figcaption{The population-level (rectangular panels) and individual-system (square panels) marginalized mutual inclination distributions for the two $N=10$ samples of fake CB disks around SB2s. The top half of the figure is devoted to the isotropic sample and the bottom half shows the low mutual inclination sample. In the rectangular subpanels, we show twenty random samples of the inferred mutual inclination distribution (blue lines) overlaid with the true distribution used to generate the sample (black lines; $p_\mathrm{iso}(\imut)$  and $p_\mathrm{low}(\imut)$, respectively). The framework can clearly differentiate generating distributions that are isotropic and nearly coplanar, without mistaking one for the other. In the square panels, we show the marginal posterior of the inferred $\imut_j$ value for each system in the sample, with the true $\imut_j$ value for that system denoted by a vertical black line. The individual disk posteriors for the isotropic sample appear two-horned from the marginalization over $\phi_j$ when $i_\mathrm{disk} \not \simeq i_\star$; i.e., in Figure~\ref{fig:disk-frame} a large teal circle (constant $\imut$) is required to intersect the red annulus (constant $i_\star$ + uncertainty), which it tends to do at two values of $\theta$. These peaks do not appear in the low inclination sample because when $i_\mathrm{disk} \simeq i_\star$ the teal circle can be small and inscribe a path entirely contained within the uncertainties of the red-annulus.
  \label{fig:fake}}
  \end{center}
\end{figure*}

For a sample of 10 systems our hierarchical model requires $32$ parameters. This high dimensional parameter space is common to hierarchical problems and is challenging to explore using MCMC algorithms popular with astronomers, such as the Metropolis-Hastings algorithm and the affine-invariant ensemble sampler \citep{goodman10,foreman-mackey13}. One approach to hierarchical sampling is the $K$-samples method \citep[as described in][]{hogg10b}, whereby independent samples from (lower-dimensional) individual disk posteriors are reweighted under the hierarchical prior to approximately calculate $p(\boldsymbol{\alpha}|\,\boldsymbol{D})$. This approach is useful when the modeler has access to samples of the likelihood for each system but not the individual datasets required to calculate the likelihood directly. Because we have access to the data of the individual systems themselves (in Table~\ref{tab:disks-known}), we choose to directly sample the full high dimensional posterior with a version of Hamiltonian Monte Carlo (HMC) algorithm called the No U-Turn Sampler \citep{hoffman14}, implemented in \texttt{PyMC3} \citep{pymc3}. 

In addition to evaluating the posterior density at each sample, HMC algorithms also calculate the gradient of the posterior with respect to the model parameters in order to simulate the evolution of a dynamic system. The gradient is calculated to machine precision using automatic differentiation as provided by \texttt{PyMC3} through the \texttt{Theano} framework \citep{theano}. The use of gradient information makes HMC samplers very efficient in higher dimensionality spaces and effective at exploring the highly correlated pathologies common to hierarchical models \citep{betancourt13,betancourt17}. Via the MCMC sampling process, we obtain joint samples of $\{\boldsymbol{\mu}_j\}_{j=1}^N$ from the full posterior distribution. It is straightforward to obtain samples of the $\{\imut_j\}_{j=1}^N$ and $\boldsymbol{\alpha}$ marginalized over the other dimensions by simply dropping these other dimensions from the multivariate output chain. 

We apply our hierarchical Bayesian model separately to each of the two samples using only the information in $\boldsymbol{D} = \{\cos i_{\mathrm{disk},j}, \cos i_{\star, j} \}_{j=1}^N$ and show marginal posteriors of the mutual inclinations in Figure~\ref{fig:fake} (top half is the ``iso'' sample, bottom half is the ``low'' sample). In the wide subpanels we show realizations of the mutual inclination distribution $p(\imut | \boldsymbol{D}, \boldsymbol{\alpha})$ generated from draws from the marginal posterior of $p(\boldsymbol{\alpha} |\, \boldsymbol{D})$. We represent the mutual inclination distribution that generated each fake dataset with a black curve. In both instances, we infer mutual inclination distributions that closely hew to the true distribution. We deliberately chose fake distributions $p_\mathrm{iso}$ and $p_\mathrm{low}$ that were not explicit subsets of our mutual inclination parameterization to demonstrate that $p(\imut|\,\boldsymbol{\alpha})$ has sufficient flexibility to accurately model these distributions, regardless. The remainder of the subpanels in Figure~\ref{fig:fake} show the marginal posteriors of $p(\imut_j | \boldsymbol{D})$ for the individual disks in each of the samples. We mark the true mutual inclination for each system with a black line. For the isotropic sample, although the inferences of the individual mutual inclinations are broad, the posterior does bracket the true $\imut_j$ in every case. For the low mutual inclination sample, we see that the mutual inclination of each system is inferred to be low. These drastically different distributions were recovered using the same hierarchical model implementation (and choice of hyperpriors) with access to only the values of $i_\mathrm{disk}$ and $i_\star$ in each sample, demonstrating that the hierarchical model has the power to accurately discriminate between a range of generating mutual inclination distributions. 

\begin{figure}[t]
\begin{center}
  \includegraphics{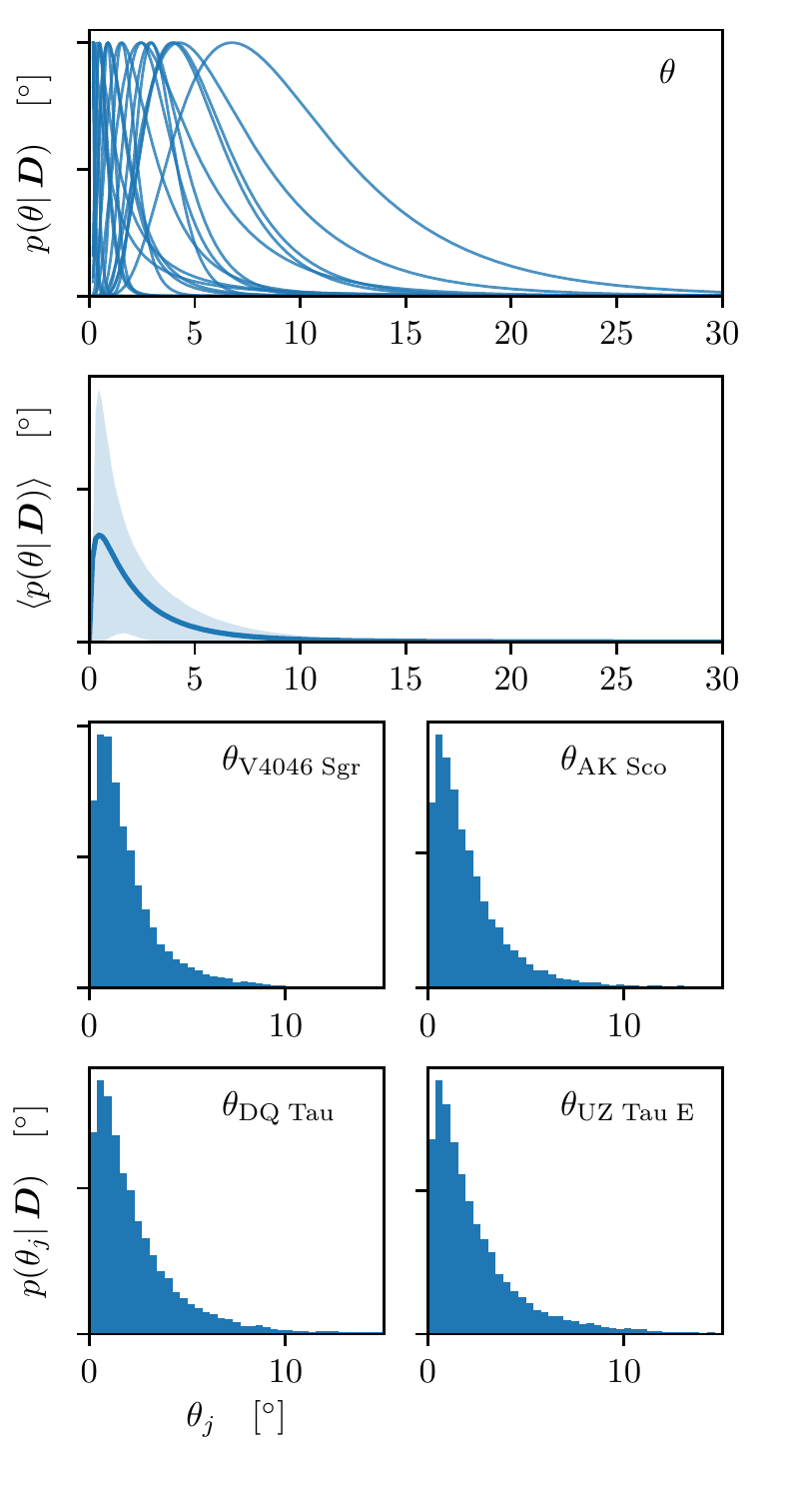}
  \figcaption{Same as Figure~\ref{fig:fake}, but utilizing the subsample of protoplanetary disks around spectroscopic binaries:  \vsgr, \ak, \dq, and \uzte. In the second panel, we show the mean value of $p(\theta |\,\boldsymbol{D})$ with a 68\% confidence interval calculated from all of the draws.
  \label{fig:real}}
  \end{center}
\end{figure}

\paragraph{Inferring the mutual inclinations of the subsample of CB disks around SB2s}
Finally, we apply our hierarchical Bayesian model to the subsample of CB disks around SB2s: \vsgr, \dq, \uzte,  and \ak. Even though there is an astrometric orbit for the \ak\ binary \citep{anthonioz15}, the model used for the interferometric fit included a narrow ring with radius $\sim$0.5\,au to mimic the contribution from the inner edge of a circumbinary disk. Subsequent scattered light observations \citep{janson16,dong16} and sub-mm observations (Czekala et al., \emph{in prep}) revealed that the dust in \ak\ is actually distributed in a narrow ring with a radius of $\sim$30\,au, raising the possibility that the stellar orbital parameters from the interferometric model could be biased. For the purposes of the hierarchical model, we ignore the \ak\ astrometric orbit and treat $\Omega_\star$ and $\imut$ as unknown. 

We present the mutual inclination distribution inferred from this subsample in Figure~\ref{fig:real}. In addition to plotting 20 random draws of the mutual inclination distribution, in the second panel we also represent $p(\imut |\,\boldsymbol{D})$ in an alternate form by showing the mean value of the $p(\imut |\,\boldsymbol{D})$ draws with a shaded 68\% confidence interval. These two representations are equivalent visualizations of the uncertainty in the inferred mutual inclination distribution. The distribution clearly favors low mutual inclinations, with 68\% of the probability contained within $\imut < 3.0\degr$ (95.4\% and 99.7\% within $\imut < 9.5\degr$ and $\imut < 11.9\degr$, respectively). We summarize the marginal $\imut_j$ posteriors of the individual disks in Table~\ref{tab:disks-known}.

\begin{figure*}[t]
\begin{center}
  \includegraphics{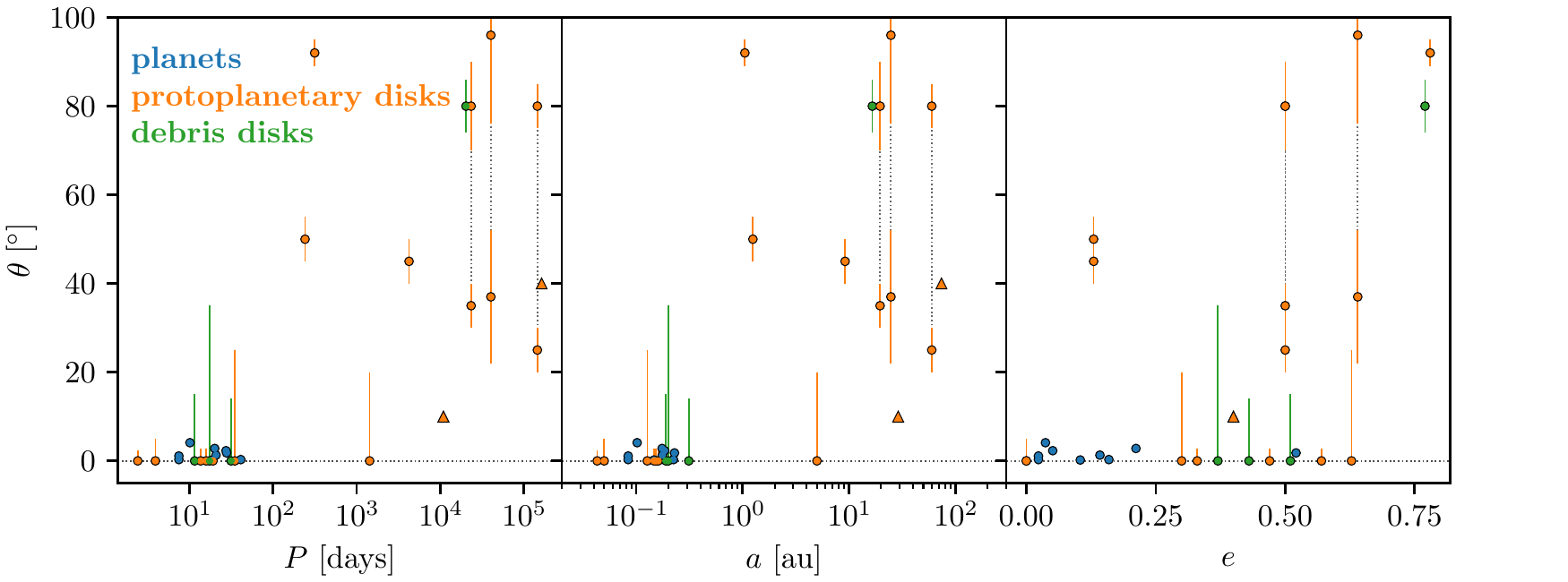}
  \figcaption{\emph{Left}: The mutual inclinations of \kepler\ circumbinary planets and all circumbinary protoplanetary and debris disks in Tables~\ref{tab:disks-known} and \ref{tab:disks-quasi}, as a function of binary orbital period. The triangles represent the \emph{lower} limits on $\imut$ for R~CrA and IRS\,43. The dotted lines connect degenerate solutions for HD~142527, SR\,24N and GG Tau Aa-Ab. \emph{Center}: Mutual inclination as a function of semi-major axis. \emph{Right}: Mutual inclination as a function of binary eccentricity. The triangle represents the \emph{lower} limit on $\imut$ for R~CrA ($e$ is unknown for IRS\,43, and so it is not plotted in the eccentricity panel). The two points at $e=0.13$ correspond to GW~Ori A-B and AB-C, which collectively host a circumternary protoplanetary disk. 
  Long-period, eccentric binaries are more likely to host circumbinary disks with significant mutual inclinations.
  \label{fig:triptych}}
  \end{center}
\end{figure*}

\subsection{Mutual inclination as a function of orbital period}
We now have a sample of 19 binary orbits with circumbinary (or circumternary) disks with mutual inclination measurements (Tables~\ref{tab:disks-known} and \ref{tab:disks-quasi}). We plot these systems, along with the \kepler\ circumbinary planets, as a function of period, semi-major axis, and eccentricity in Figure~\ref{fig:triptych}. We find that of the six protoplanetary CB disks orbiting stars with $P < 40$\,days, five of them have mutual inclinations constrained to $\imut <5\degr$; the sixth (TWA~3A) is consistent with this sample, though with a larger uncertainty on $\imut$. All three of the debris disks in this period range are also consistent with having low mutual inclinations, although their upper limits on $\imut$ are also not as stringent ($\imut \lesssim 25\degr$). At these shorter binary periods, the distribution of CB disk mutual inclinations is consistent with that of the Kepler CB planets, which all have $\imut < 5\degr$ and orbit binaries with $P \leq$\,41\,days. 

For the ten CB disks orbiting binaries with periods longer than 40 days, the distribution of mutual inclinations differs dramatically, with most systems having values of $\imut > 20\degr$. Although the sample size is limited, it appears as though there are two clusters of systems, substantially misaligned systems with $\imut \sim 40\degr$ and drastically misaligned systems with near-polar disk orientations $\imut \sim 90\degr$. Only one disk (HD~200775) has a mutual inclination $\imut < 20\degr$. There is a significant trend that misaligned disks surround only the most eccentric binaries ($e \gtrsim 0.3$; see the last paragraph of section \ref{subsec:disk-binary} for a discussion of the point at $(e,\theta) = (0.1, 45^\circ)$ corresponding to the GW~Ori circumternary disk). 

It is unlikely that these trends result from observational selection effects. The probability of detecting a disk from its spatially unresolved thermal radiation is independent of disk inclination, provided that radiation originates from material that is optically thin (which debris disks are at all wavelengths, and protoplanetary disks may be in the sub-mm). In addition, we know of no observational bias that would correlate (in whatever sense) the orientations of binary orbits with those of their surrounding disks. For example, while radial velocity searches for spectroscopic binaries are more sensitive to edge-on orbits, and while debris disks may be more easily spatially resolved when viewed edge-on owing to higher line-of-sight column densities, there is no bias that would correlate their relative nodal orientations on the sky.

\section{Discussion}\label{sec:discussion}

\subsection{Disk-Binary Interactions} \label{subsec:disk-binary}
Turbulent fragmentation of a collapsing molecular core \citep{offner10} and fragmentation of a protostellar disk by gravitational instability \citep{kratter08} are two ways by which stellar binaries can directly form. However, neither of these mechanisms appears capable of forming protostars with initial separations $a < 10$\,au. Direct fragmentation on small scales requires high densities which renders gas optically thick, supported by thermal pressure, and resistant to cooling and collapse \citep{larson69,bate02}.

Close binaries are likely made instead from initially wide binaries that underwent subsequent hardening \citep[e.g.,][and references therein]{bate19}. A variety of mechanisms exist to shrink binary orbits: dynamical interactions with third parties (via, e.g., Lidov-Kozai oscillations and tidal friction), dissipative interactions with circumstellar and circum-multiple disks, and the accretion of low angular momentum gas onto the binary from a residual core/envelope \citep{artymowicz96,bate02,bate12}. \citet{moe18a} found that Lidov-Kozai cycles are insufficient to explain the full population of close binaries: even among those binaries hosting a tertiary star, 60\% require extra dissipative interactions with primordial gas to reach their current separations. In a large radiation-hydrodynamical simulation of a collapsing molecular cloud, \citet{bate19} found that a close binary typically forms when two initially unbound stars become bound and shrink their orbit by gravitational interactions mediated by their disks. Gas-rich disks provide both a larger cross-section for encounters and a means of dissipating orbital energy. 

Binaries that form by dissipative encounters may be orbited by disks having an initially wide range of mutual inclinations \citep{bate18}. The inclinations can evolve by subsequent disk-binary interactions. \citet{foucart13,foucart14} found that when the binary orbit is circular or nearly so, and when the relative disk-binary inclination is small (but non-zero), gravitational torques between the warped disk and the binary can bring their mean planes into alignment. These authors computed an alignment timescale that is short compared to the disk lifetime unless the viscosity governing inclination damping is much smaller than the viscosity controlling radial diffusion of mass (see also \citealt{lodato13}), or the inner edge of the disk is far removed from the binary.

If, however, the binary is substantially eccentric, it can force the disk out of alignment. The orbit of a circumbinary test particle, if initially sufficiently inclined, librates (oscillates) about $\theta = 90^\circ$;\footnote{The fixed point at $\theta = 90^\circ$ represents the strongest,
quadrupole-order resonance. Other resonances with other fixed points are surveyed by \citet{vinson18} in the test particle limit.} the more eccentric the binary, the smaller is the initial inclination required to access this libration \citep{verrier09,farago10,naoz17,vinson18}. These librations are damped in a viscous disk, which seeks to settle into a polar configuration \citep{martin17,zanazzi18,lubow18} or near-polar configuration if the disk mass is high \citep{martin19b}. In the parameter survey of \citet{martin18}, disks initially misaligned by $30^\circ$ or more around a binary with $e=0.8$ undergo damped librations about $\theta = 90^\circ$; disks with smaller initial inclinations are not attracted to the polar configuration but still exhibit inclination variations of tens of degrees (see, e.g., their Figure 9).

As documented in Tables~\ref{tab:disks-known} and \ref{tab:disks-quasi}, binaries with $P < 10 \,{\rm days}$ are circularized---presumably from tidal dissipation. In support of the theoretical ideas discussed above, Figure \ref{fig:triptych} attests that CB disks orbiting short-period binaries are co-planar (or nearly so) with their hosts, while CB disks orbiting long-period, eccentric binaries exhibit a variety of mutual inclinations ranging up to $\theta \approx 90^\circ$. The polar orientations of HD~98800B and 99~Her are consistent with damped librations about the fixed point of $\theta = 90^\circ$ in eccentric binaries \citep{martin17}; these disks also exhibit the nodal orientations expected from this scenario \citep[e.g.,][]{zanazzi18,lubow18,kennedy19}.\footnote{By contrast, those polar orbits that fall within the observational uncertainties for HD~142527, SR~24N, and GG~Tau~A do not evince the predicted nodal orientation.} Those disks that are not polar or substantially inclined around binaries with eccentricities of $\sim$0.5 may have formed with inclinations below the threshold required for polar librations \citep{foucart14,martin18}. A few of these disks orbit binaries with periods of 10--40 days, and may have had their inclinations damped while driving their hosts into more compact configurations. Notably no CB disk is obviously retrograde,\footnote{Technically there are observational ambiguities which permit retrograde solutions for HD~200775, R~CrA, and IRS~43.} presumably because the dissipative disk-star encounters that most effectively bind binaries do not involve retrograde motions \citep{borkovits16,tokovinin17,bate18}.

The two points in the right panel of Figure~\ref{fig:triptych} at $(e,\theta) \simeq (0.13, 45^\circ)$ and $(0.13, 50^\circ)$ pertain to the disk orbiting the GW~Ori hierarchical triple. The orbital
eccentricities of the triple, which itself is co-planar, 
appear too low for the mechanism of \citet{martin18} to generate
the observed disk inclination. The large $\theta$ might
instead be a relic of the original dissipative disk-mediated
encounter that presumably formed the multiple system \citep{bate18}.
Alternatively, if the GW Ori disk holds the
bulk of the system's angular momentum,
Lidov-Kozai oscillations may have shaped the system.

\subsection{Circumbinary Planets} \label{subsec:cbp}
Our finding that CB disks around short-period, spectroscopic binaries are nearly co-planar ($\imut < 3.0\degr$) adds to the various lines of evidence that \kepler's CB planets---so far discovered with periods shorter than $\sim$300 days---are similarly co-planar with their host stars.  \citet{armstrong14} found that by assuming either a strictly co-planar CB planet population ($\imut = 0\degr$) or a nearly co-planar distribution centered around $\imut = 5\degr$ (similar to our $p_\mathrm{low}$ in \S\ref{subsec:hbm}), the CB planet occurrence rate is either $7_{-1}^{+5}\%$ or $13_{-2}^{+10}\%$, respectively, for planets with radii $\in [4, 10] R_\earth$. Their assumed $\theta$-distributions resemble the one we have measured for CB disks around short-period binaries, and deliver occurrence rates consistent with that around single stars for similar planet radii and period ranges \citep[$\sim $8\%;][]{fressin13}. The interpretation that the \kepler\ CB planets have low mutual inclinations is also supported by the Bayesian analysis of \citet{li16}, who used the effects of orbital stability and finite observing time to break the degeneracy between planet occurrence rate and mutual inclination and concluded that CB planets must on average have $\imut < 3\degr$. The BEPOP radial velocity search for CB planets around single-lined eclipsing binaries also claims $\imut \lesssim 10 \degr$ \citep{martin19}.

An enduring curiosity is that eclipsing binaries (EBs) having periods $< 7$ days have not been found to host CB planets, despite the fact that most known \kepler\ EBs (whether planet-hosting or not) have shorter periods and thus higher transit probabilities for correspondingly shorter period, dynamically stable planets \citep{armstrong14,li16}. Lidov-Kozai oscillations driven by a tertiary companion have been invoked to tilt planets out of transit or crash them onto their host stars or be ejected \citep{munoz15,martin15b}, but as previously noted the role of Lidov-Kozai in forming compact binaries seems limited at best \citep{moe18a,kounkel19}. Indeed the two CB disks in our sample around $P < 7$\,day binaries, \vsgr\ and CoRoT~2239, have small mutual inclinations and do not host tertiaries capable of driving significant Lidov-Kozai oscillations \citep[\vsgr\ hosts a companion orbiting $\sim$12,000\,au outside the disk;][]{kastner11}. Two promising alternative explanations for the shortfall of CB planets around the shortest-period binaries are 1) a primordial phase of binary orbit expansion driven by pre-main-sequence tidal evolution which destabilizes planetary orbits \citep{fleming18} and 2) increased X-ray and EUV flux from tight, tidally locked binaries photoevaporates circumbinary planets to smaller planet radii \citep{sanz-forcada14}, which would remain undetected in the \kepler\ light curves.

\section{Conclusions}
\label{sec:conclusion}
We have analyzed new ALMA observations of the circumbinary (CB) protoplanetary disk around the double-lined spectroscopic binary (SB2) \uzte, finding individual stellar masses of $M_\mathrm{Ea} = 0.93 \pm 0.04 \,M_\odot$ and $M_\mathrm{Eb} = 0.28 \pm 0.02 \,M_\odot$, and a sky-projected inclination of the binary similar to that of the disk ($i_\star = 56.1\pm5.7\degr$, $i_\mathrm{disk} = 56.2 \pm 1.5$). \uzte\ joins a sample of three other CB disks around short-period SB2s, all having $i_\mathrm{disk} \simeq i_\star$ and binary periods $P < 20$\,days. Although the striking similarity of sky-referenced disk and binary inclinations in these short-period systems suggests that true disk-binary mutual inclinations $\imut$ are small, technically we cannot calculate $\imut$ definitively because the binaries and their disks have unknown relative nodal orientations. We have circumvented this difficulty by implementing a hierarchical Bayesian analysis that fully leverages the observation that $i_\mathrm{disk} \simeq i_\star$ to infer, in a statistical sense, that short-period binaries are indeed nearly co-planar with their surrounding disks, with 68\% of systems having $\imut < 3.0\degr$. 

We have assimilated the above sample into a larger collection of the best-characterized circumbinary protoplanetary and debris disks orbiting binaries with periods ranging up to $\sim$$10^5$ days. Many of these systems have astrometric measurements of the stellar orbit, which enables their mutual inclinations to be calculated directly. Disk-binary mutual inclinations are found to trend strongly with both binary period $P$ and binary eccentricity $e$: all CB disks orbiting binaries with $P<30$\,days and/or $e < 0.2$ are consistent with being co-planar, while CB disks orbiting longer period and/or more eccentric binaries exhibit a wide range of mutual inclinations, from co-planar to unambiguously polar (HD~98800B and 99~Her).

These trends are consistent with our current understanding of close binary star formation and gravitational torques exerted between host binaries and dissipative CB disks. Binary stars with semi-major axes $< 10\,$au are thought to form at larger separations and be possibly initially unbound; subsequent dissipative disk-binary gravitational interactions reduce the total system energy and bind/harden the orbit. During these early times, circumbinary disk gas, predominantly prograde, may be initially variously inclined with respect to the binary \citep{bate14,bate18}. Those binaries on circular orbits---which the shortest period binaries tend to be as a result of tidal friction---may drive their CB disks into co-planar alignment, assuming initially modest mutual inclinations \citep{foucart13,foucart14}. Short-period binaries may correlate with initially small disk misalignments insofar as orbital migration driven by CB disks is more effective at small $\imut$. Eccentric binaries, on the other hand, can induce large disk misalignments. If the initial $\theta$ exceeds a minimum threshold (whose value decreases as the binary eccentricity increases; for $e = 0.8$ the minimum $\imut$ is about $30\degr$), then $\theta$ can be driven to its equilibrium value of $\approx$$90\degr$ \citep{martin17,martin18,zanazzi18,lubow18}. Even if the initial $\theta$ falls short of the critical value, the non-axisymmetric potential of the eccentric binary can secularly force the CB disk to change its inclination by tens of degrees \citep{martin18,vinson18}.

Our finding that CB disks around short period binaries are nearly co-planar with their stellar hosts implies (at face value) that planets spawned by such CB disks should be similarly co-planar. We thus add to the growing evidence that \kepler\ CB planets, orbiting binaries with $P < 40$\,days, have small mutual inclinations, and that by extension the CB planet occurrence rate (in this parameter space) is similar to that for single stars \citep{armstrong14,li16,martin19}. Beyond $P > 40$ days, however, the existence of both aligned and misaligned CB disks leads us to expect that CB planets around long-period, eccentric binaries will be discovered to have a correspondingly broad distribution of mutual inclinations, with a possible concentration of systems having reached their evolutionary endpoint at $\theta \approx 90^\circ$.

\acknowledgments
\emph{Acknowledgements} ---
IC thanks Rob De Rosa, Eric Nielsen, and Lea Hirsch for many helpful discussions about astrometric orbit conventions; Gaspard Duch\^ene for discussions related to stellar multiplicity; Dan Foreman-Mackey for discussions about gradient-based inference frameworks; David Martin and Daniel Fabrycky for discussions about circumbinary planets; J.J. Zanazzi and Dong Lai for discussions about polar alignment mechanisms; Maxwell Moe for discussions about stellar multiplicity and Lidov-Kozai oscillations, and Jiaqing Bi for discussions about GW Ori. 
IC and EC thank Steve Lubow for discussions. 
IC was supported by NASA through the NASA Hubble Fellowship grant HST-HF2-51405.001-A awarded by the Space Telescope Science Institute, which is operated by the Association of Universities for Research in Astronomy, Inc., for NASA, under contract NAS5-26555. This paper makes use of the following ALMA data: ADS/JAO.ALMA\#2015.1.00690.S. ALMA is a partnership of ESO (representing its member states), NSF (USA) and NINS (Japan), together with NRC (Canada), MOST and ASIAA (Taiwan), and KASI (Republic of Korea), in cooperation with the Republic of Chile. The Joint ALMA Observatory is operated by ESO, AUI/NRAO and NAOJ. 
The National Radio Astronomy Observatory is a facility of the National Science Foundation operated under cooperative agreement by Associated Universities, Inc.
This research has made use of the NASA/IPAC Extragalactic Database (NED), which is operated by the Jet Propulsion Laboratory, California Institute of Technology, under contract with the National Aeronautics and Space Administration. This research has made use of the SIMBAD database, operated at CDS, Strasbourg, France. This research has made use of NASA's Astrophysics Data System Bibliographic Services.

\software{CASA \citep[v4.4;][]{mcmullin07}, DiskJockey \citep{czekala15a,disk_jockey_zenodo}, RADMC-3D \citep{dullemond12}, emcee \citep{foreman-mackey13}, Astropy \citep{astropy13}, PyMC3 \citep{pymc3}, Theano \citep{theano}, Daft (PGM)}.

\bibliographystyle{yahapj.bst}
\bibliography{circumbinary.bib}

\appendix

\begin{figure*}[ht!]
\begin{center}
  \includegraphics{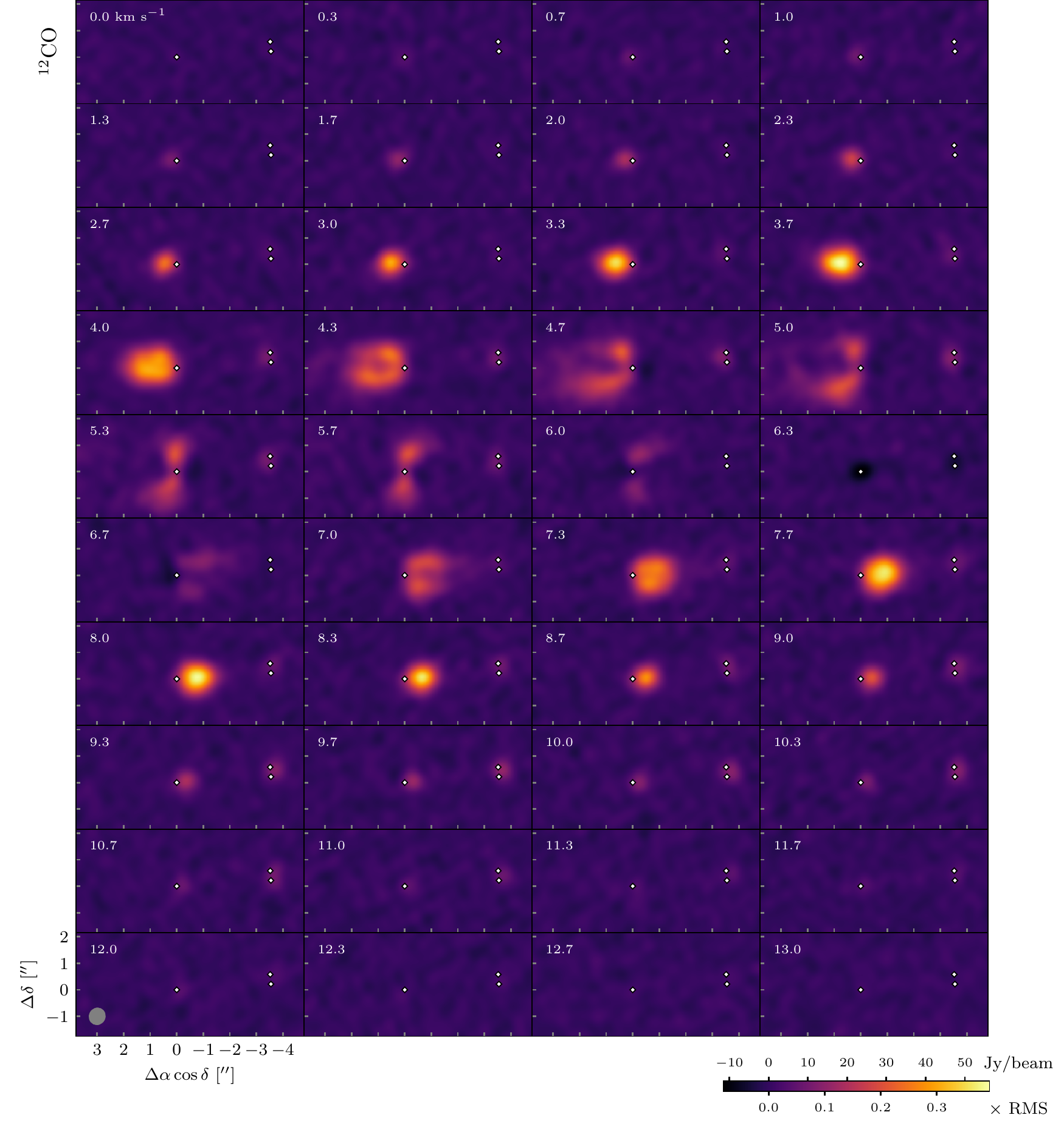}
  \figcaption{Channel maps for \twelve. The positions of \uzte, Wa, and Wb are marked as in Figure~\ref{fig:moments}. All velocities are in the LSRK frame.
  \label{fig:chmaps-12}}
  \end{center}
\end{figure*}

\begin{figure*}[ht!]
\begin{center}
  \includegraphics{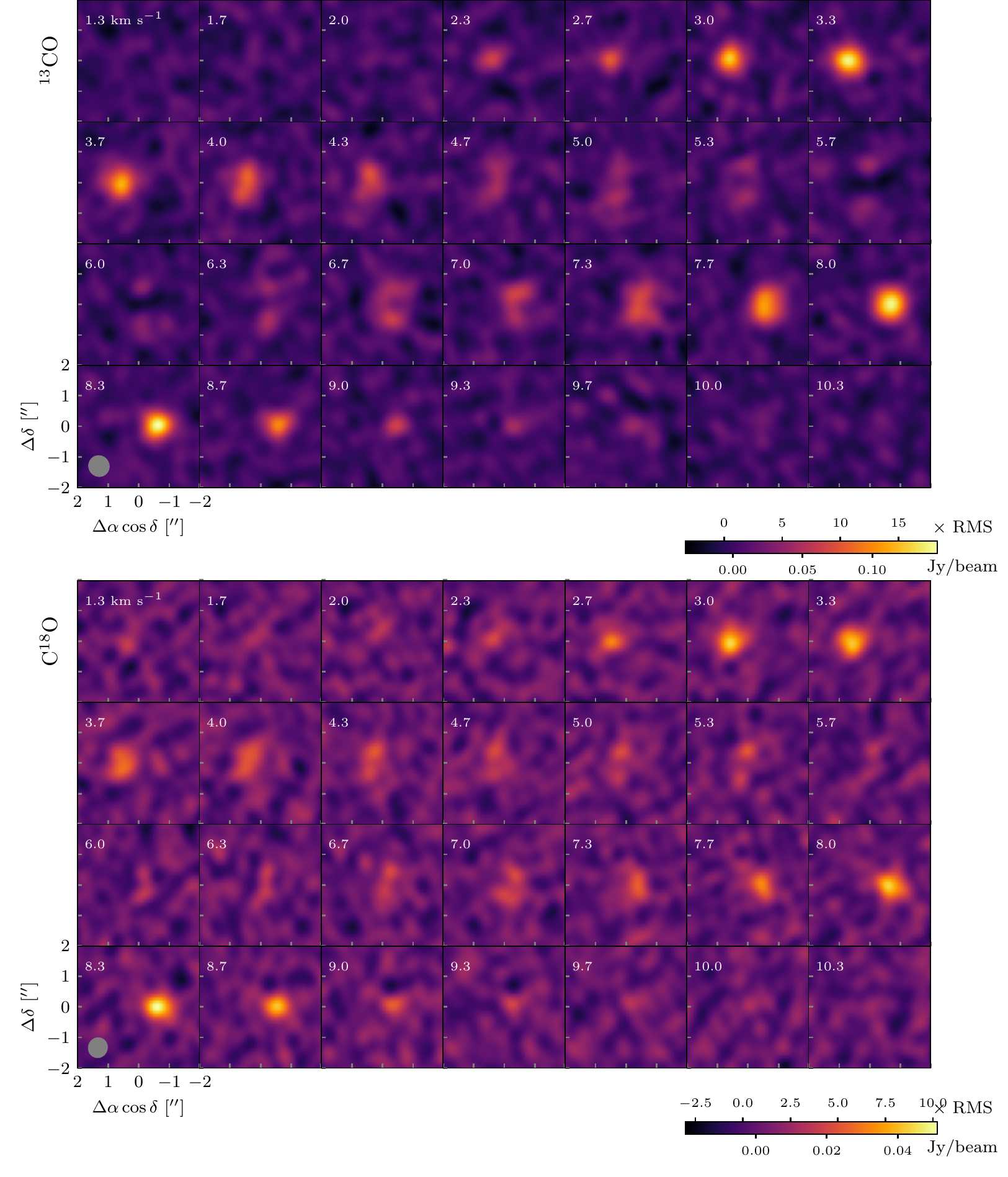}
  \figcaption{Channel maps for \thirteen\ and \eighteen\ isotopologues. The maps are shown centered on \uzte\ because the emission from \uztw\ is not evident in individual frames. All velocities are in the LSRK frame.
  \label{fig:chmaps-13-18}}
  \end{center}
\end{figure*}

\begin{figure*}[ht!]
\begin{center}
  \includegraphics{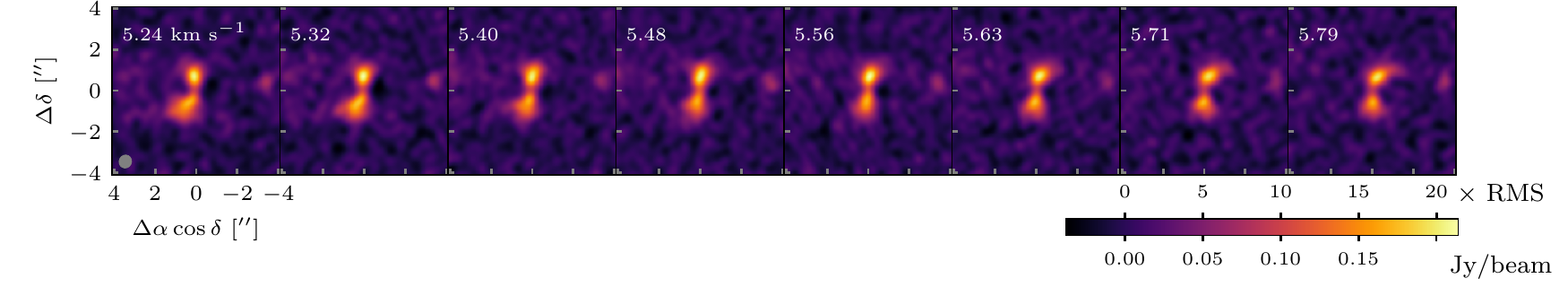}
  \figcaption{\twelve\ channel maps centered on \uzte\ at the highest effective spectral resolution of $79\,\rm m\thinspace s^{-1}$. All velocities are in the LSRK frame. Based upon the location of the symmetric ``figure-$\mathsf{8}$'' lobe, we determine the systemic velocity of \uzte\ to be $v_\mathrm{LSRK} = 5.5 \pm 0.1\,\kms$, or $v_\mathrm{bary} = 15.5 \pm 0.1\,\kms$.
  \label{fig:chmaps-high}}
  \end{center}
\end{figure*}

\begin{figure*}[ht!]
\begin{center}
  \includegraphics{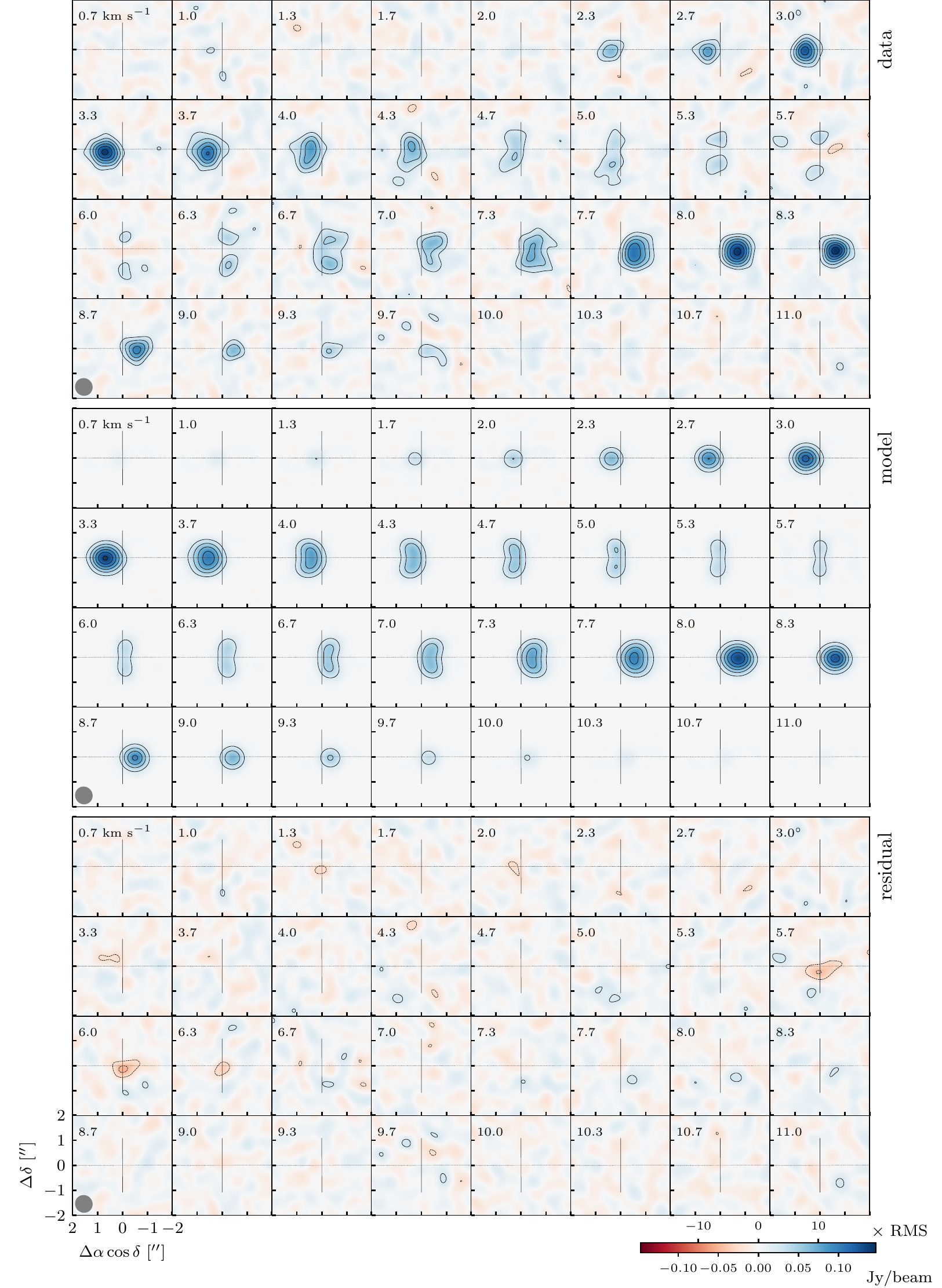}
  \figcaption{\thirteen\ channel maps centered on \uzte. All velocities are in the LSRK frame. Contours are in multiples of three times the RMS. Channels in the range $4.0 \leq v_\mathrm{LSRK} \leq 7.5\,\kms$ were excluded from the fit to avoid cloud contamination.
  \label{fig:chmaps-13-resid}}
  \end{center}
\end{figure*}

\begin{figure*}[ht!]
\begin{center}
  \includegraphics{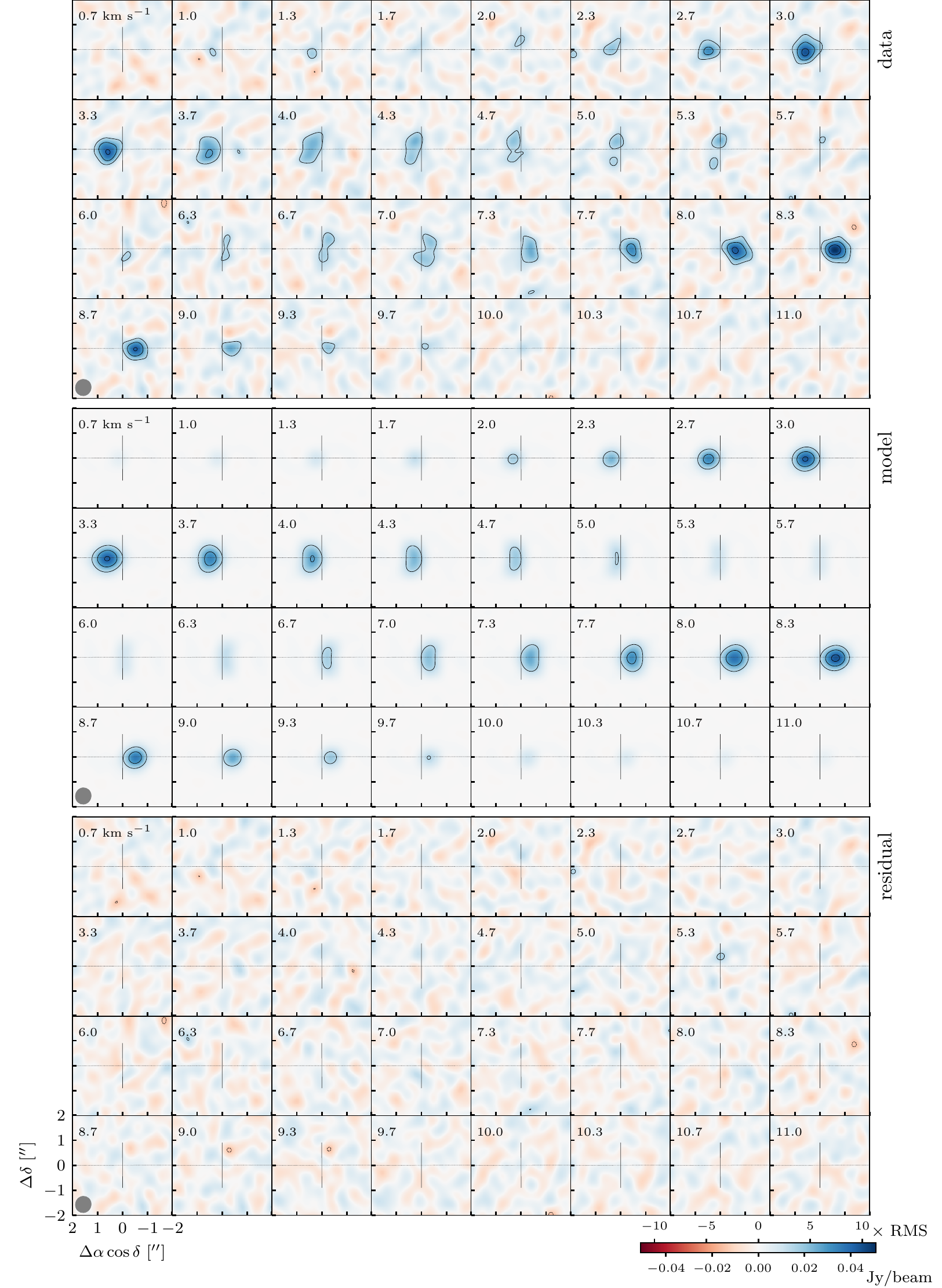}
  \figcaption{\eighteen\ channel maps centered on \uzte. All velocities are in the LSRK frame. Contours are in multiples of three times the RMS. Channels in the range $4.0 \leq v_\mathrm{LSRK} \leq 7.5\,\kms$ were excluded from the fit to avoid cloud contamination.
  \label{fig:chmaps-18-resid}}
  \end{center}
\end{figure*}

\end{document}